\documentclass[12pt,preprint]{aastex}

\shorttitle{NLTE line-formation for hydrogen revisited}
\shortauthors{Przybilla \& Butler}

\begin{document}

\title{Non-LTE line-formation for hydrogen revisited}

\author{Norbert Przybilla\altaffilmark{1}}
\affil{Institute for Astronomy, 2680 Woodlawn Drive, Honolulu, HI 96822, USA}
\altaffiltext{1}{present address: Dr. Remeis-Sternwarte
Bamberg, Sternwartstra{\ss}e 7, D-96049 Bamberg, Germany}
\email{przybilla@sternwarte.uni-erlangen.de}

\and

\author{Keith Butler}
\affil{Universit\"atssternwarte M\"unchen, Scheinerstra{\ss}e 1, D-81679 M\"unchen, Germany}
\email{butler@usm.uni-muenchen.de}

\begin{abstract}
We discuss aspects of non-LTE line formation for hydrogen in early-type stars. 
We evaluate the effect of variations in the electron-impact excitation cross
sections in model atoms of differing complexity by comparison with observation.
While the Balmer lines are basically unaffected by the choice of atomic data, the 
Paschen, Brackett and Pfund series members allow us to discriminate between the
different models. Non-LTE calculations based on the widely-used
approximation formulae of Mihalas, Heasley \& Auer and of Johnson
fail to simultaneously
reproduce the optical and IR spectra over the entire parameter range. The
use of data from {\em ab-initio} calculations up to principal quantum number 
$n$\,$\leq$\,7 largely solves the problem.
We recommend a reference model using the available data. This model is of
general interest because of the ubiquity of the hydrogen spectrum.
\end{abstract}


\section{Introduction}
The quantitative interpretation of the hydrogen line spectrum is one of the
foundations of modern astrophysics. Being the most abundant and most
basic element in the universe hydrogen imprints its signature on the
spectra of the majority of astronomical objects. The analysis of these line
features allows us to determine the physical properties of stars, nebulae
and accretion phenomena. For decades the focus was on the first members of
the Balmer series,
easily accessible from the ground, with many studies concentrating on
understanding, verifying and improving the\,modelling of\,these key\,indicators. 

In the meantime developments in detector technology have opened the infrared
(IR) window to routine observation, in some bands from the ground, in
its entirety from space. IR observations will gain in importance in the
future as the next generation of ground-based large telescopes
and the next large space telescope will also be operating in this wavelength
range. This development is driven mainly by a change of focus to the
high-$z$ universe, but it will also allow local objects in 
otherwise inaccessible environments to be studied, e.g.\ ultra-compact
\ion{H}{2} regions, the Galactic centre and dust-enshrouded nearby starburst galaxies.
The Brackett and Pfund lines are important diagnostics
at these wavelengths. It is natural to ask whether the present modelling 
is in similar good shape for their quantitative interpretation as it is for
the Balmer lines. 

The present work therefore addresses several aspects of non-LTE line formation for
hydrogen in early-type stars. The next two sections are dedicated to finding a reference
\ion{H}{1} model atom in order to provide a diagnostic for both visual and IR
spectra. This is done by testing different implementations
by comparing with observation. 
The resulting set of reference data is, of course, of much broader interest than for
stellar analyses alone. Relevant data from new {\em ab-initio}
computations for electron-impact excitation of hydrogen are presented in an
appendix.

\section{Model calculations}

\subsection{Model atmospheres and programs}
The line-formation computations are carried out using two methods. For
main sequence stars of spectral types later than O and BA-type supergiants a
hybrid approach is chosen. Based on hydrostatic, plane-parallel, 
line-blanketed LTE models calculated with the {\sc Atlas9} code (Kurucz~1993)
the non-LTE computations are performed using {\sc Detail} and {\sc Surface}
(Giddings~1981; Butler \& Giddings~1985). The coupled radiative transfer and 
statistical equilibrium equations are solved with the former and the emergent 
flux computed with the latter. This line-formation package has undergone
significant modifications recently, most notably through the inclusion of an
ALI scheme (using the treatment of Rybicki \& Hummer~1991). Line-blocking
is realised by considering Kurucz' ODFs (Opacity Distribution Functions). 

For the modelling of early B and O-type stars we use the non-LTE 
model-atmosphere/ line-formation code {\sc Fastwind} (Santolaya-Rey, Puls \&
Herrero~1997) which accounts for spherical extension and hydrodynamic
mass-outflow. It has been recently updated to include an approximate treatment of non-LTE
line-blocking/blanketing (Puls et al.~2003).

\subsection{Atomic data}
The two-body nature of the hydrogen atom allows the radiative data to be
obtained analytically. On the other hand,
excitation and ionization processes involving a colliding particle
require a numerical solution of the resulting three-body Coulomb problem.
A number of quantum-mechanical {\em ab-initio} calculations exist for excitation
via electron collisions which reproduce the few measurements for transitions
from the ground state fairly well. However, for transitions from excited
states one has to rely on theory. For the majority of
the transitions only approximation formulae are available. These should
provide data accurate to a factor better than two, some claiming the 
uncertainties to be as small as $\sim$20\%. Similar restrictions are found 
for electron-impact ionization data, which however are less important in practice.

Our model atoms comprise levels with principal quantum number up to
$n$\,$\leq$\,50, and energies adopted from Moore \cite{Moore93}. All lines
are included assuming Stark profiles for transitions between energy levels
with $n$\,$\leq$\,7, applying the theory of Griem~\cite{Griem60} as implemented 
by Auer \& Mihalas~\cite{AuMi72}, and Doppler profiles for the remainder.
Transition probabilities are calculated using a routine of Storey \&
Hummer~\cite{StoHum91}. Photoionization cross-sections and the 
free-free opacity are evaluated applying hydrogenic expressions 
(Mihalas~1978, p.\,99 \& p.\,102) with Gaunt factors treated according to
Wright~\cite{Wright78} (see Appendix B).

\begin{figure}
\includegraphics[angle=-90,width={0.49\columnwidth}]{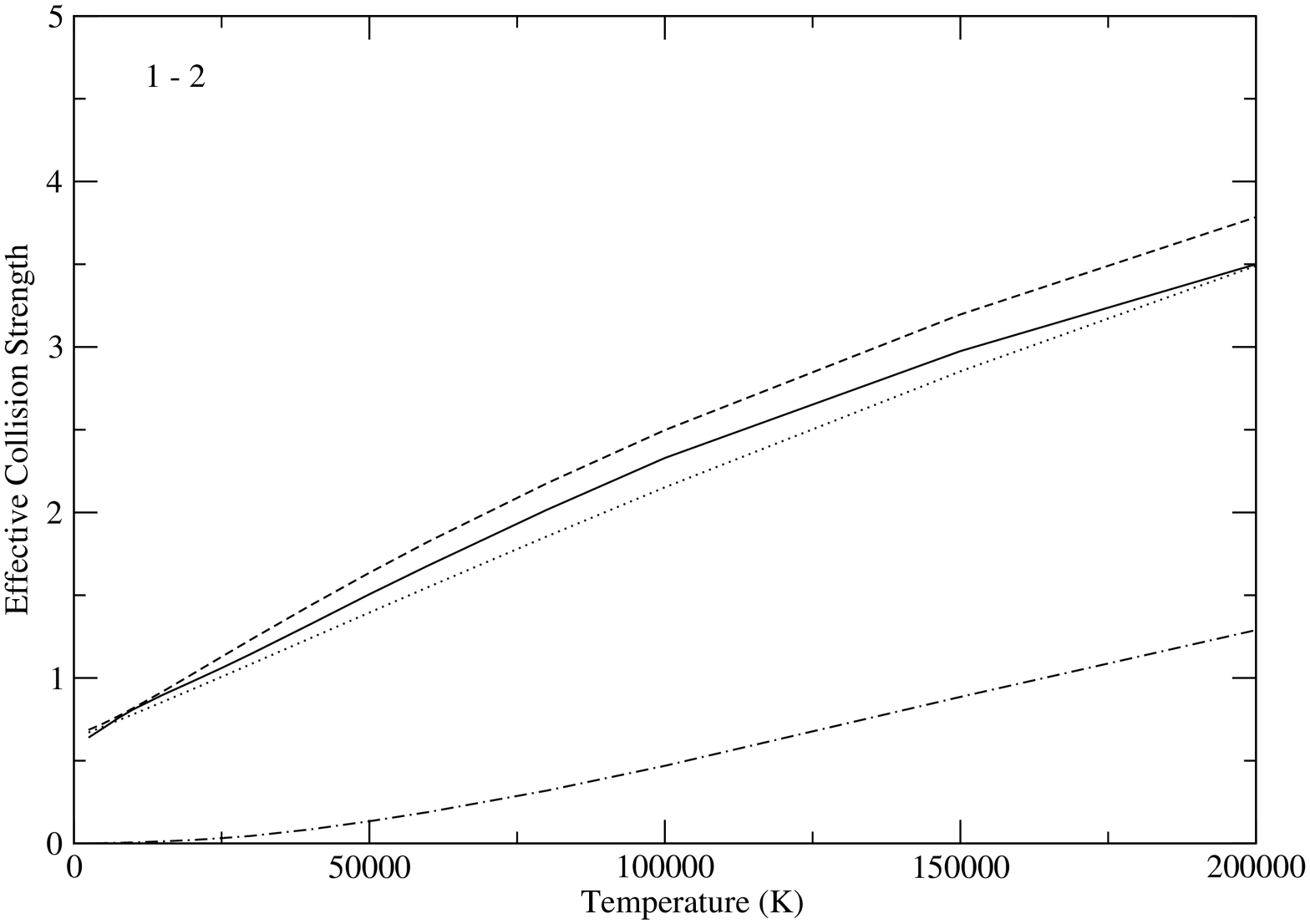}\\[-.6cm]
\includegraphics[angle=-90,width={0.49\columnwidth}]{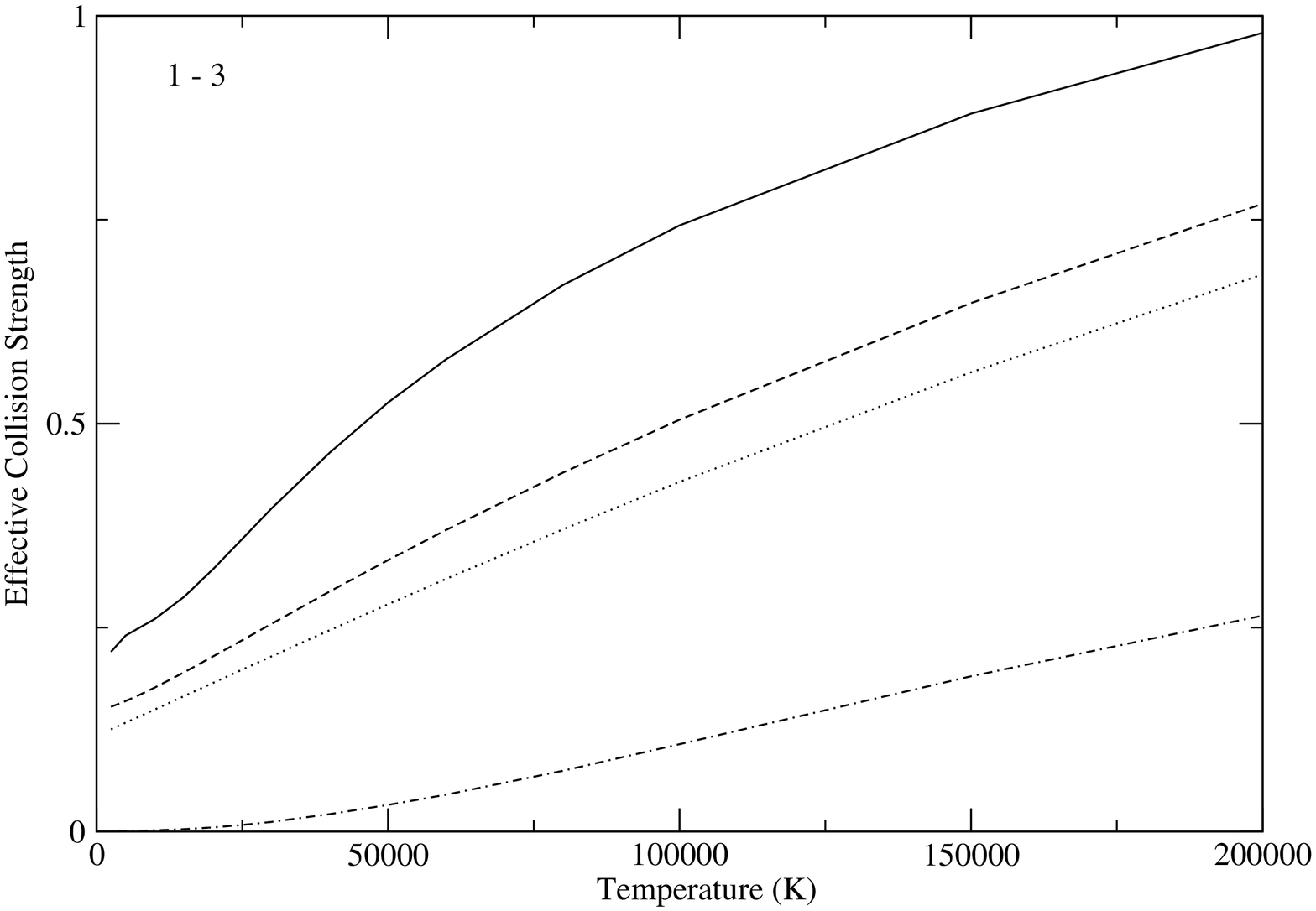}\\[-.6cm]
\includegraphics[angle=-90,width={0.49\columnwidth}]{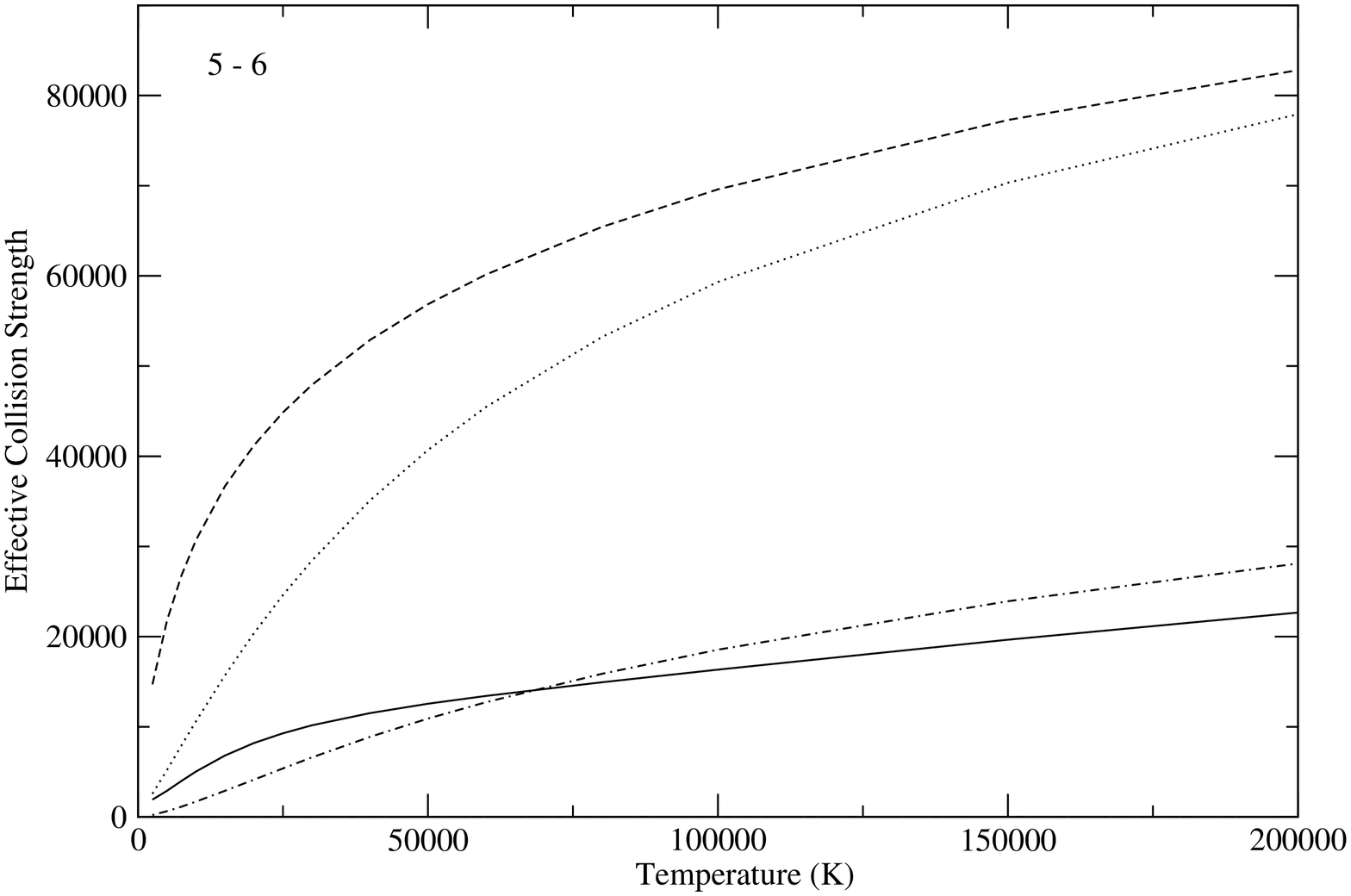}
\figcaption{Comparison of effective collision strengths for several
transitions $n$\,--\,$n$', as indicated. The curves are: B04 ({\em
solid}), J72 ({\em dotted}), MHA ({\em dashed}), PR ({\em dash-dotted}).
\label{coll_1_2}}
\end{figure}

We have constructed model atoms using data for electron-impact excitation 
according to Mihalas, Heasley \& Auer~\citep[MHA]{MHA75}, the approximation 
formulae of Johnson~\citep[J72]{Johnson72} and 
Percival \& Richards~\citep[PR]{PeRi78} 
and the {\em ab-initio} calculations of Ander\-son et
al.~\cite[ABBS]{Andersonetal00, Andersonetal02} and Butler (in preparation, B04)
for transitions between lower/upper levels $n$,$n'$, see Table~\ref{tabmod}.
A few of the collision rates are illustrated in Fig.~\ref{coll_1_2}.  
All the rates are
in good agreement for the $n$\,$=$\,1--2 transition since accurate data were
available to Johnson and MHA in this case.  The classical results of
PR are only valid for $n$, $n$$^\prime$ greater or equal to 5.  They are only
included in this case for reference.  On the other hand there is considerable
disagreement for the $n$\,$=$\,1--3 and $n$\,$=$\,5--6 transitions although in 
the latter case
the PR results are close to those of B04. Overall both the Johnson and MHA
data show equally large departures from the modern results.  For the higher
$n$ values the agreement between PR and B04 is better but there are still
differences.  These may be due to inadequacies in either approach and the
question as to which data are to be preferred for these high n transitions
can only be answered by more extensive calculations.
In addition, the numerical fits of Giovanardi, Natta \&
Palla~~\citep[GNP]{GNP87} have been applied to the collisional
excitation rates in some test cases. 
Electron-impact ionization rates are evaluated according to
Mihalas~\cite{Mihalas67} for $n$\,$\leq$\,10 and by use of the
Seaton formula~\cite{Seaton62} for $n$\,$>$\,10. For test purposes the
approximation formula of Johnson~\cite{Johnson72} has also been implemented for 
collisional bound-free processes, without significant impact.

In the final step with {\sc Surface} the hydrogen spectrum is synthesised
adopting wavelengths from Wiese, Smith \& Glennon~\cite{Wieseetal66}. Stark
broadening of the Balmer and Paschen lines is accounted for by the use of
the tables of Stehl\'e \& Hutcheon~\citep[SH]{SH99}. In the case of the
Brackett and Pfund lines we apply the theory of Griem~\cite{Griem60} as 
implemented by Auer \& Mihalas~\cite{AuMi72}.

\subsection{Model complexity}
In addition to the choice of the atomic data, the complexity of the model atom
can also influence the model predictions. In order to investigate this
effect, model atoms are constructed with 10 to 50 explicitly treated energy 
levels. The model atoms are identified by a label 
plus the number of levels in the following. Both departure coefficients and 
line source functions obey an asymptotic behaviour (see below), such that the 
tests help to identify the minimum requirements for model\,complexity. 

A complication arises as a result of the interacton of the hydrogen atoms with the
stellar plasma. High-lying energy levels are broadened as they are
perturbed, and finally dissolved -- the Stark broadening of the hydrogen lines
near the series limits gives rise to a quasi-continuum. Our straightforward approach 
reproduces the observed series limit behaviour (highest-frequency series
member with separately discernible profile, i.e. the classical Inglis-Teller
limit, and the flux distribution throughout the line region and quasi-continuum) 
in the cases where the available spectra cover these regions. Despite this
success, the solution is not fully self-consistent, and future work will
incorporate the occupation probability formalism of Hubeny, Hummer \&
Lanz~\cite{HHL94} for improvement.

\subsection{Discussion}
The mechanisms driving departures of \ion{H}{1} from detailed balance in
stellar atmospheres have been well understood since the seminal work of Auer \&
Mihalas~\citep[for early-type stars]{AuMi69a,AuMi69b}, and numerous
subsequent contributions -- for line formation in the IR e.g. by Zaal et
al.~\cite{Zaaletal99}. We do not intend to repeat these arguments. We only
wish to point out that the atomic data on radiative processes in \ion{H}{1} 
are of high accuracy, thus excluding a source of systematic uncertainty
compromising studies of most ions of the other elements. 
The issue here is the impact of the {\em local} processes that affect the
radiatively induced departures
from LTE, namely collisional interactions, which are assumed to be of secondary
importance. Indeed, various choices of the (mostly approximate)
data produce no significant differences
in the stellar continuum or the Balmer line profiles, i.e. the features that
are the starting point for quantitative analyses using model atmosphere
techniques. 

\begin{figure}
\plotone{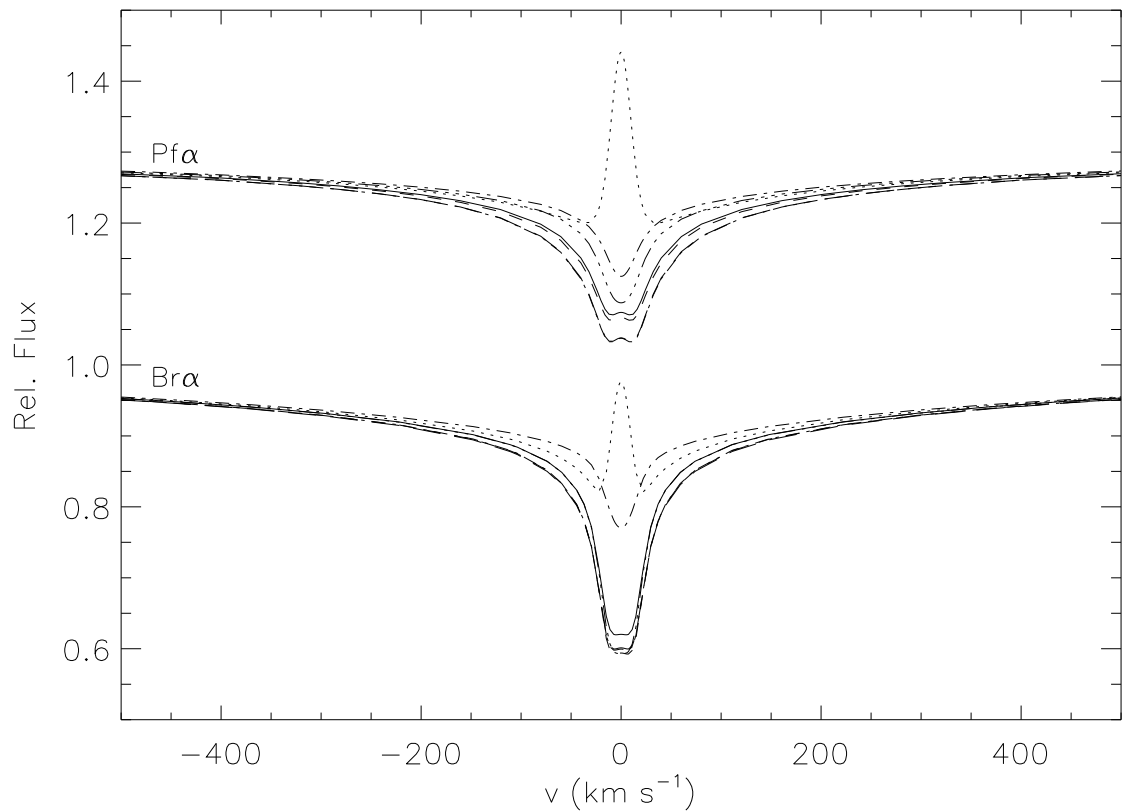}
\figcaption{Comparison of model profiles for Br$\alpha$ and Pf$\alpha$ in
Vega: models A, C, E, F ({\em dotted, dash-dot-dot-dotted, full, dashed
lines}) and LTE ({\em dashed-dotted}). Models B, C and D are not resolved  
in the case of Br$\alpha$ as they practically coincide with model F,
forming the lowest set 
of graphs, as do models B and D in the case of Pf$\alpha$.
The non-LTE computations are for a 20-level model atom. 
Abscissa is Doppler velocity relative to line centre. For
clarity the different spectra are shifted by 0.3 units in ordinate.
\label{vega_alpha}}
\end{figure}

But consider Fig.~\ref{vega_alpha}, where we summarise the
results from our model calculations for Br$\alpha$ and Pf$\alpha$ for
atmospheric parameters that match those of
Vega, one of the best-studied stars.
Apparently, the choice of collisional data is not a second-order
effect, but a dominant factor for
line-formation computations in the IR. The only way to determine which data
should be used is by comparison of the models with observation over all 
parameter space. 

Before this is done in the following section, we wish to discuss the effects of 
the different atomic model implementations in more detail, in
anticipation of the results to come. We choose our model
for $\beta$\,Ori 
to illustrate this, as the non-LTE
effects are amplified in a supergiant (although the r\^ole of collisions
are somewhat diminished in such tenuous atmospheres), but the results are similar 
in the general stellar context. They also allow more general conclusions to
be drawn, as the atomic data are valid -- under
normal circumstances -- independent of environment.

\begin{figure}
\plotone{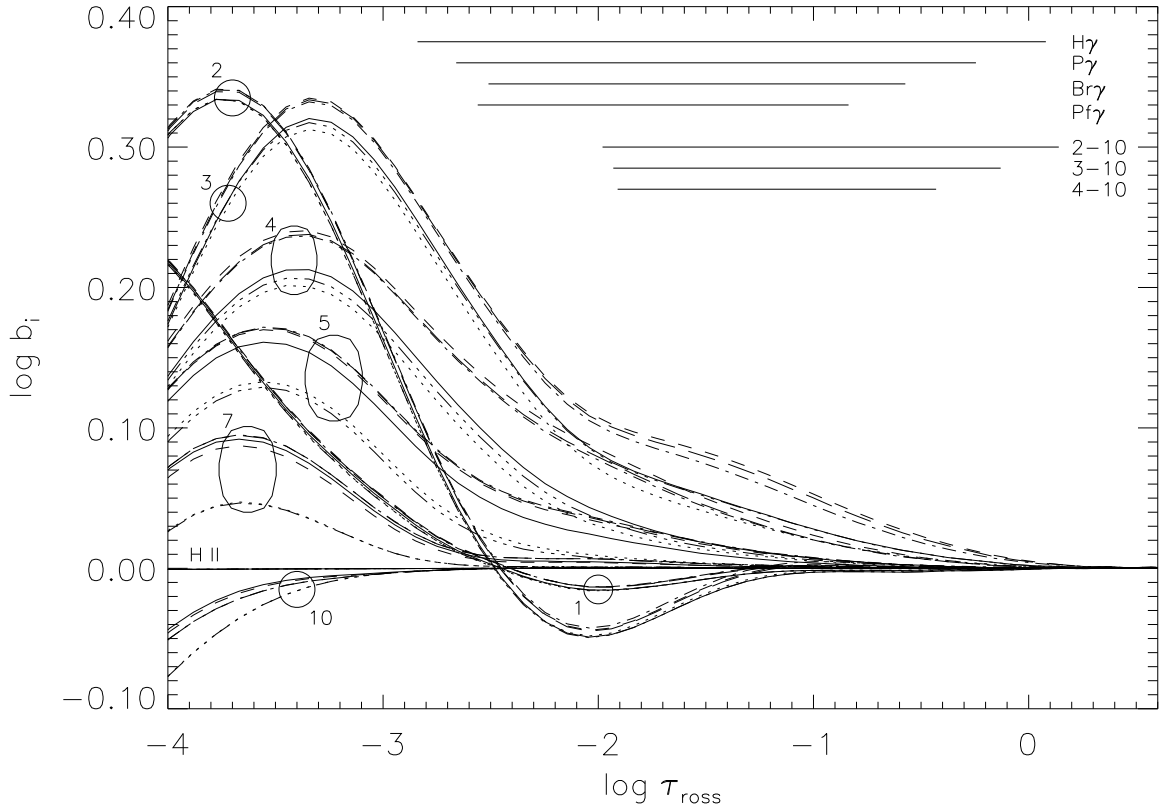}
\figcaption{Run of departure coefficients $b_i$ in $\beta$\,Ori as a
function of Rosseland optical depth $\tau_{\rm ross}$,
models A--F ({\em dotted, dashed-dotted, dash-dot-dot-dotted, long dashed,
full, dashed lines}). All non-LTE computations are for a 30-level model atom.
The individual sets of graphs are labelled according to the level's principal
quantum number; all graphs for \ion{H}{2} coincide. Line-formation depths for 
a few features are indicated.
\label{bori_b30}}
\end{figure}

\begin{figure}
\plotone{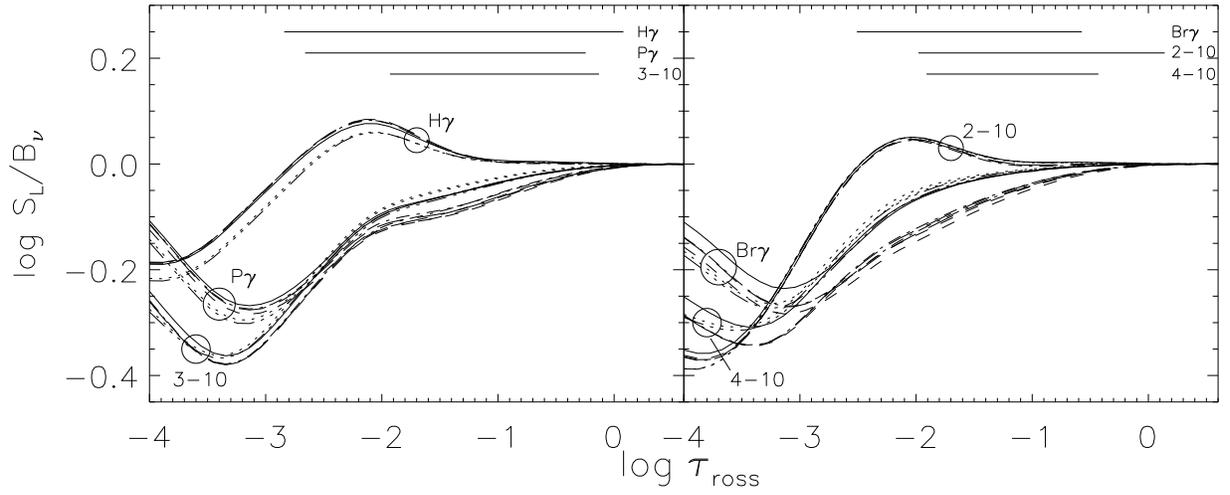}
\figcaption{Ratio of line source function $S_{\rm L}$ to Planck function
$B_{\nu}$ at line centre as a function of $\tau_{\rm ross}$ in $\beta$\,Ori.
Line designations as in Fig.~\ref{bori_b30}.
\label{bori_soverb}}
\end{figure}

Departure coefficients $b_i$\,$=$\,$n_i/n_i^{\ast}$ (the $n_i$ and
$n_i^{\ast}$ being the non-LTE and LTE populations of level $i$, respectively) 
for selected levels using models A30--F30 are displayed in
Fig.~\ref{bori_b30}. The overall behaviour, i.e. the over- and
underpopulation of the levels of the minor ionic species and the major
species \ion{H}{2}, is governed by the radiative 
processes, while the differences in the collisional data lead to modulations.
These are small for the ground state and become only slightly more pronounced
for the $n$\,$=$\,2 level, as these are separated by comparatively large
energy gaps from the remainder of the term structure. Only colliding particles 
in the high-velocity tail of the
Maxwellian velocity-distribution are able to overcome these energy
differences at the temperatures encountered in the star's atmosphere. 
Thus, computations of the model atmosphere structure will not be
significantly influenced, as the important bound-free opacities of hydrogen
vary only slightly. Line-formation computations in the IR on the other hand
will be affected, as maximum effects from variations of the collisional data
are found for the levels with intermediate $n$ at line-formation depth.
In fact, the differences in the collisional
cross-sections from approximation formulae and {\em ab-initio}
computations are largest for transitions among the $n$\,$=$\,3--7 levels 
with $\Delta n$\,$=$\,1 and 2, and they can amount to more than an order of
magnitude. The higher Rydberg states again show less sensitivity
as they approach the limiting case of LTE, which is independent of the details of
individual (de-)populating mechanisms. Detailed collisional cross-sections from 
{\em ab-initio} computations up to $n$\,$\simeq$\,7 are therefore sufficient
to eliminate a significant source of systematic error.
However, using the available data it turns out that the
MHA- (models A, C, E) and J72-type approximations (models B, D, F) give rise to 
basically two 
sets of distinct behaviour, with the former tending to dampen non-LTE departures more 
efficiently than the latter, due to larger collisional cross-sections.
Such differences in the level populations are the cause for the
line-profile variations. This is also evident in Fig.~\ref{bori_soverb},
where the response of the line-source function $S_{\rm L}$ to variations in
the collisional data is displayed. Here, models A and F usually define both
extremes. The IR lines experience a non-LTE strengthening in this star. 
Comparison with observed IR lines will be highly
valuable in selecting the preferable collisional data.
We remember that the line source function may be written as
\begin{equation}
S_l = \frac{2h\nu^3/c^2}{b_i/b_j\exp(h\nu/kT) - 1}
\end{equation}
so that the source function is particularly sensitive to 
variations in the ratio of the departure coefficients
\begin{eqnarray}
\left|\Delta S_l\right| &=&
\left| \frac{S_l}{b_i/b_j -\exp(-h\nu/kT)}\Delta (b_i/b_j)\right|\cr
&\approx& \left| \frac{S_l}{(b_i/b_j -1)+h\nu/kT}\Delta (b_i/b_j)\right|
\label{eqn2}
\end{eqnarray}
when $h\nu/kT$ is small. This makes these lines very susceptible to small
changes in the atomic data and details of the calculation. Indeed, the
emission cores of Br$\alpha$ and Pf$\alpha$ predicted by model A in Vega
(Fig.~\ref{vega_alpha}) result from such a non-LTE amplification. For
Pf$\alpha$ the $\Delta (b_5/b_6)$ between model A and E amounts to only
$\lesssim$2\% at the line-core formation depth, with the $b_5/b_6$ ratios
for the different model atoms deviating even less from unity.

Note that because of our restricted approach (plane-parallel, hydrostatic
atmosphere) this discussion is only realistic for the inner photosphere
($\log \tau_{\rm ross}$\,$\gtrsim$\,$-$2) in this supergiant, so that significant 
differences between the model predictions and observations for the first members 
of each series can be expected with regard to their line-formation depths.
For H$\gamma$ this typically manifests itself in the line core where the
observed profile is shallower than that predicted, while the line wings are
reproduced. In the more luminous supergiants H$\alpha$ develops into a completely
wind-dominated P-Cygni line. The modelling of the higher series members on the 
other hand can be expected to be more accurate, as they are formed deeper in
the photosphere.   

\begin{figure}
\plotone{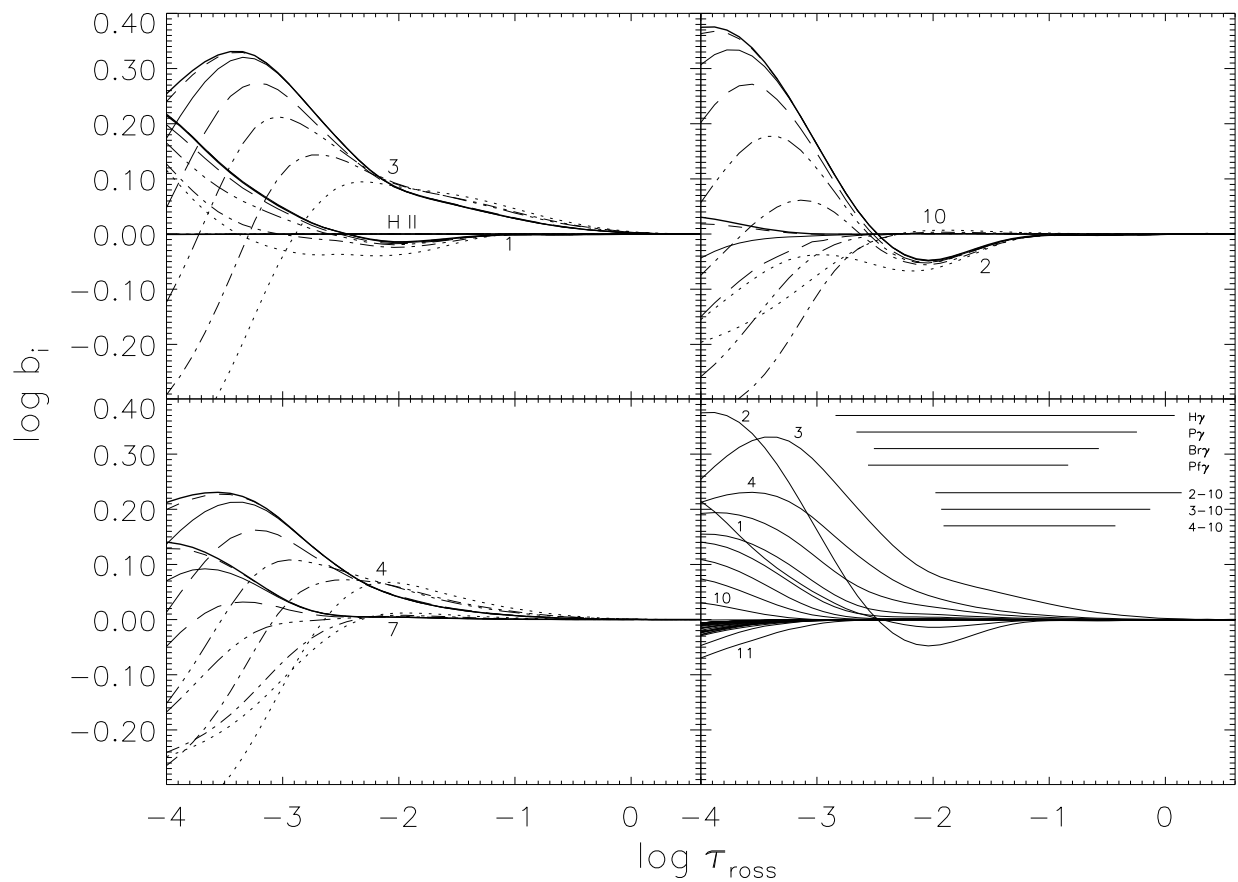}
\figcaption{Run of $b_i$ in $\beta$\,Ori as a function of $\tau_{\rm ross}$
for models E of different complexity: a 10, 15, 20, 25, 30, 40, 50-level model
({\em dotted, dashed-dotted, dash-dot-dot-dotted, long dashed,
full, dashed, full thick lines}). 
The lower right panel
displays the behaviour of all model E50 levels. Note the Rydberg states
asymptotically approaching the \ion{H}{2} limit. 
\label{bori_bE}}
\end{figure}

\begin{figure}
\plotone{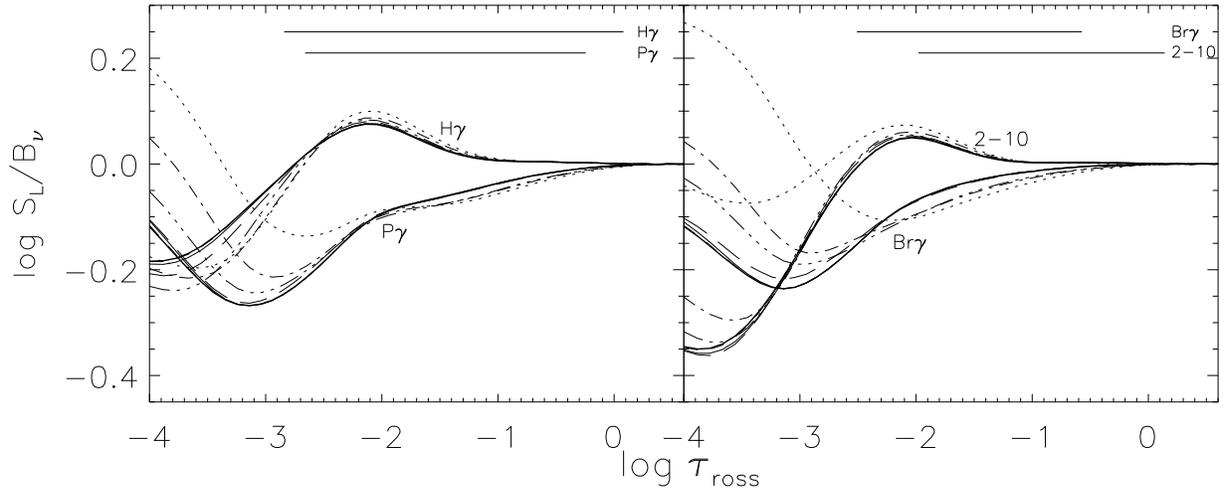}
\figcaption{As Fig.~\ref{bori_soverb}, but for models E of different
complexity (10--50 levels). Line designations as in Fig.~\ref{bori_bE}.
\label{bori_soverbE}}
\end{figure}

As indicated in the last section, another issue is of concern, the
complexity of the model atom, if level
dissolution is not explicitly accounted for.
Level populations may vary with the number of explicitly
treated states, as more channels for the (de-)population of individual levels
are opened.  The results from test calculations for our model of
$\beta$\,Ori are shown in 
Figs.~\ref{bori_bE} and \ref{bori_soverbE}. An obvious feature is the convergent 
behaviour of departure coefficients and line source functions with increasing 
number of levels. Both asymptotically approach values set by models of
considerable complexity. In the case of this star a $\sim$25--30-level 
model atom is required to achieve this at depths relevant to line formation.
This coincides with the classical Inglis-Teller limit. More simplistic model
atoms turn out to overestimate the $n$\,$=$\,3--5 populations
at line-formation depths in particular, with the effect becoming smaller for higher $n$.
Comparison with observations of Paschen, Brackett and Pfund lines is
therefore most promising in order to verify this finding empirically.
Increasing the number of explicitly treated levels even further has only an impact 
in the outermost layers, which are not properly computed for supergiants in our 
approach. Moreover, these high-lying Rydberg states will be subject to level 
dissolution in a real plasma. However, such complex model atoms are
instructive as they nicely demonstrate how the departure coefficients of the higher
\ion{H}{1} Rydberg states approach the limit set by \ion{H}{2}, see
Fig.~\ref{bori_bE} (lower right panel), due to tight collisional coupling
i.e. the limiting case of LTE. 

Despite being instructive in their own, the previous considerations can 
only provide indirect evidence for the choice an optimum model atom. A real selection
can only happen via confrontation with observation, as approximations
still have to be relied upon to a certain degree. This is done in the
following on the basis of high-quality spectra for a few well-studied objects, 
in order to minimise the impact of other systematic error, such as uncertain 
stellar parameters.

\section{Confrontation with observation}\label{obs}

\subsection{The spectra}
This work has to rely on a variety of sources for the observations. Some of
the spectra in the visual and near-IR have been investigated in our earlier
publications.
The Vega (\objectname{HD172167}) spectrum was obtained using FOCES on the 
Calar Alto\,2.2m
telescope, and the $\eta$\,Leo (\objectname{HD87737}) and $\beta$\,Ori
(\objectname{HD34085}) 
spectra were acquired with
FEROS on the ESO\,1.5m telescope at La Silla. 
Both instruments are Echelle
spectrographs, and the data were obtained at a resolving power of
$R$\,$=$\,$\lambda/\Delta\lambda$\,$\sim$\,40000 and 48000, respectively.
More detailed information on the observations and data processing 
can be found in 
Przybilla et al.~\cite{Przybillaetal01} and Przybilla \& Butler~\cite{PrBu01}.
Medium-resolution ($R$\,$\sim$\,3000) IR spectra in the $J$, $H$ and $K$ band for 
Vega and $\eta$\,Leo are adopted from the catalogues of 
Wallace et al.~\cite{Wallaceetal00},
Meyer et al.~\cite{Meyeretal98} and Wallace \& Hinkle~\cite{WaHi97}, see
these for details. We have rectified these KPNO\,4m/FTS spectra for our
purposes. Similar data for $\beta$\,Ori have been kindly provided by A.~Fullerton
(private communication), obtained with CFHT/FTS at $R$\,$\sim$\,10000, see
Fullerton \& Najarro~\cite{FuNa98} for details.  

The blue visual spectrum for $\tau$\,Sco (\objectname{HD149438}) is adopted from 
the work of Kilian
\& Nissen~\cite{KiNi89}, who used the CASPEC spectrograph on the ESO\,3.6m
telescope ($R$\,$\sim$20--25000). Both, the H$\alpha$ and the IR spectra for
this object were kindly provided by P.A.~Zaal. The MUSICOS spectrograph on
the INT at La Palma was used to obtain the former ($R$\,$\sim$\,30000) and
UKIRT/CGS4 for the latter ($R$\,$\sim$\,14--16000), see Zaal et
al.~\cite{Zaaletal99} for details on the observations and data reduction.
Optical spectra of HD\,93250 as observed with ESO NTT/EMMI ($R$\,$\sim$\,5--6000) 
are adopted from Puls et al.~\cite{Pulsetal96}, while the $K$ band
data (CTIO\,4m/OSIRIS, $R$\,$\sim$\,1100) for this star are taken from the
2$\mu$ atlas of hot, luminous stars by Hanson, Conti \&
Rieke~\cite{Hansonetal96}. See these publications for additional
information on the observations and data reduction procedures.

For the comparison of the synthetic with observed spectra, the former are
convolved with a Gaussian of full width at half-maximum appropriate to the
instrumental resolution in all cases where this is a non-negligible factor. 
All observed spectra are radial velocity corrected to the rest frame via 
cross-correlation with appropriate synthetic spectra.

\subsection{Atmospheric parameters}
The stellar parameters adopted for model calculations for the objects discussed 
here are summarised in Table~\ref{tabpara}: spectral type, effective
temperature $T_{\rm eff}$, (logarithmic) surface gravity $\log g$, helium
abundance (by number) $y$, stellar radius $R$, (logarithmic) stellar luminosity
$\log L$, (micro-)turbulent velocity $v_{\rm t}$ and projected rotational
velocity $v \sin i$. Additional data are provided where necessary for
calculations with {\sc Fastwind}: mass-loss rate $\dot M$, wind terminal
velocity $v_{\infty}$ and wind velocity parameter $\beta$. 
All data 
have been determined from spectroscopic indicators using standard techniques,
the fundamental atmospheric parameters surface gravity and effective
temperature e.g. from fitting the Balmer line wings and several
non-LTE ionization equilibria: \ion{He}{1}/\ion{\rule{-1mm}{0mm}}{2} and/or
\ion{Si}{3}/\ion{\rule{-1mm}{0mm}}{4} for the objects with 
$T_{\rm eff}$\,$\ge$\,30000\,K, and \ion{N}{1}/\ion{\rule{-1mm}{0mm}}{2}
and \ion{Mg}{1}/\ion{\rule{-1mm}{0mm}}{2} for the cooler objects --
HD\,93250: Repolust, Puls \& Herrero~\cite{Repolustetal03}; $\tau$\,Sco: Kilian et
al.~\cite{Kilianetal91}; $\eta$\,Leo and $\beta$\,Ori: Przybilla \&
Butler~\cite{PrBu01}, Przybilla et al.~\cite{Przybillaetal01}.

\subsection{Line profile fits}
\begin{figure}
\plotone{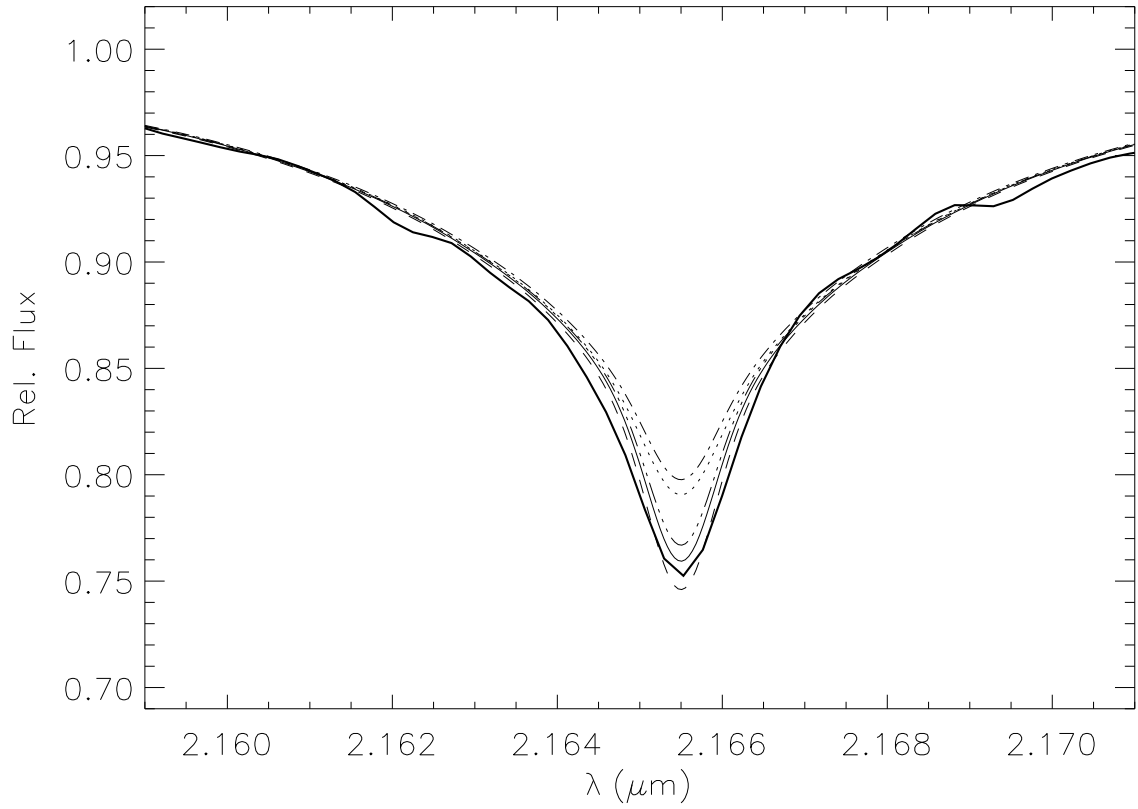}
\figcaption{Spectrum synthesis for Br$\gamma$ in Vega:
models A, C, E and F ({\em dotted, dash-dot-dot-dotted, full, dashed
lines}) and an LTE profile ({\em dashed-dotted}) are compared with observation ({\em
thick full line}). Models B and D are omitted for clarity as they are
almost identical to model F. All non-LTE computations are for a 20-level model~atom.
\label{vegabrg1}}
\end{figure}

\begin{figure}
\plotone{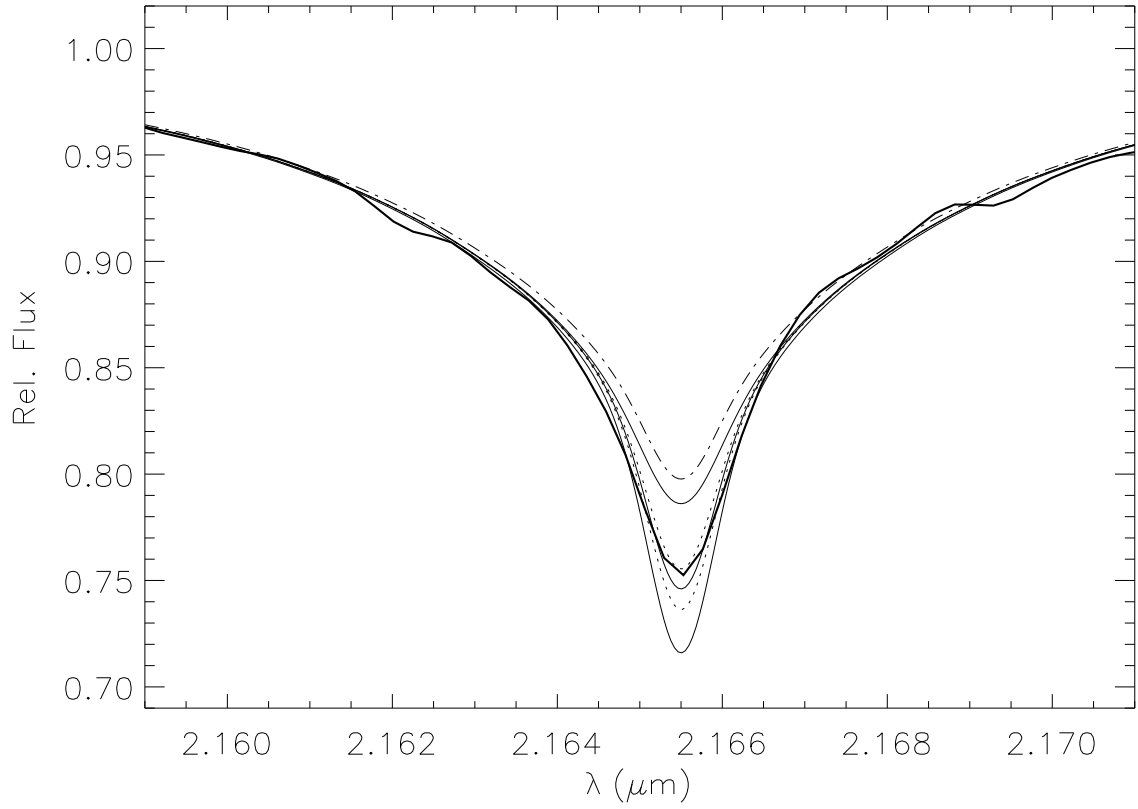}
\figcaption{Like Fig.~\ref{vegabrg1}, for a 10,
15, 20, 25 and 30-level model F ({\em full} and {\em dotted lines} in
alteration, top to bottom) and LTE ({\em dashed-dotted}).
\label{vegabrg2}}
\end{figure}

We begin our investigations with Vega, one of the most-intensely studied
stars, and in most respects well described by the assumption of LTE.
The Balmer and Paschen lines are in general well matched by LTE computations, 
and these are practically identical to the non-LTE results.
However, comparison with the observed Br$\gamma$ profile in
Fig.~\ref{vegabrg1} shows that LTE modelling fails to reproduce the line
core correctly. Non-LTE computations can improve on this, except
for model A, which differs only slightly from LTE. However, on the basis of
this one case alone a decision as to which model atom should be favoured cannot
be drawn, as despite being noticeable the differences between models B to F 
are not highly significant. Moreover, the model complexity will also affect the
line-formation computations as shown in Fig.~\ref{vegabrg2}. In this case
the 10-level model atoms would predict too shallow a line core, close to the
LTE result, and the 25- and 30-level models a line core that is too deep. Note that this 
sensitivity distinguishes model atoms using the J72 approximation from those
applying the MHA approximation, which are far less affected. 

Additional observations of Br$\alpha$ and Pf$\gamma$ in the $L$-band are
available (Wallace \& Hinkle~2002). However the lower S/N of the data
restricts their usefulness for our purposes, except for ruling out LTE modelling and 
model atom A. 
Additional complications arise in the case of Br$\alpha$, where the 
computations predict a significantly too deep line core for models B--F. This may 
be an indication of neglected non-LTE effects on the atmospheric structure, as
the line core is formed at atmospheric layers unsampled by the higher series
members. A similar effect is found in the very line core of H$\alpha$
(here the models indicate a too shallow line), which is formed at similar
atmospheric depths.

\begin{figure}
\epsscale{.78}
\plotone{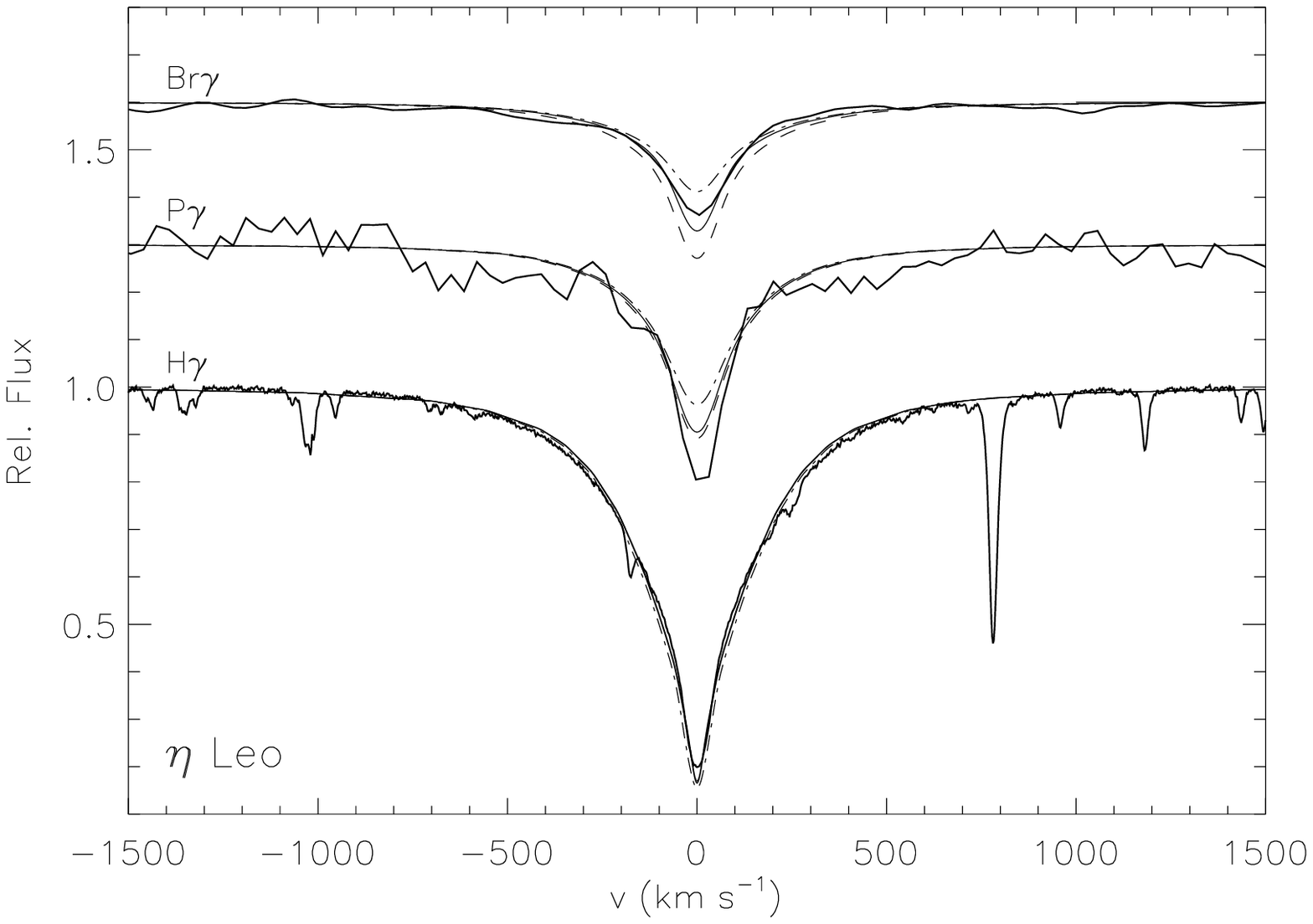}
\plotone{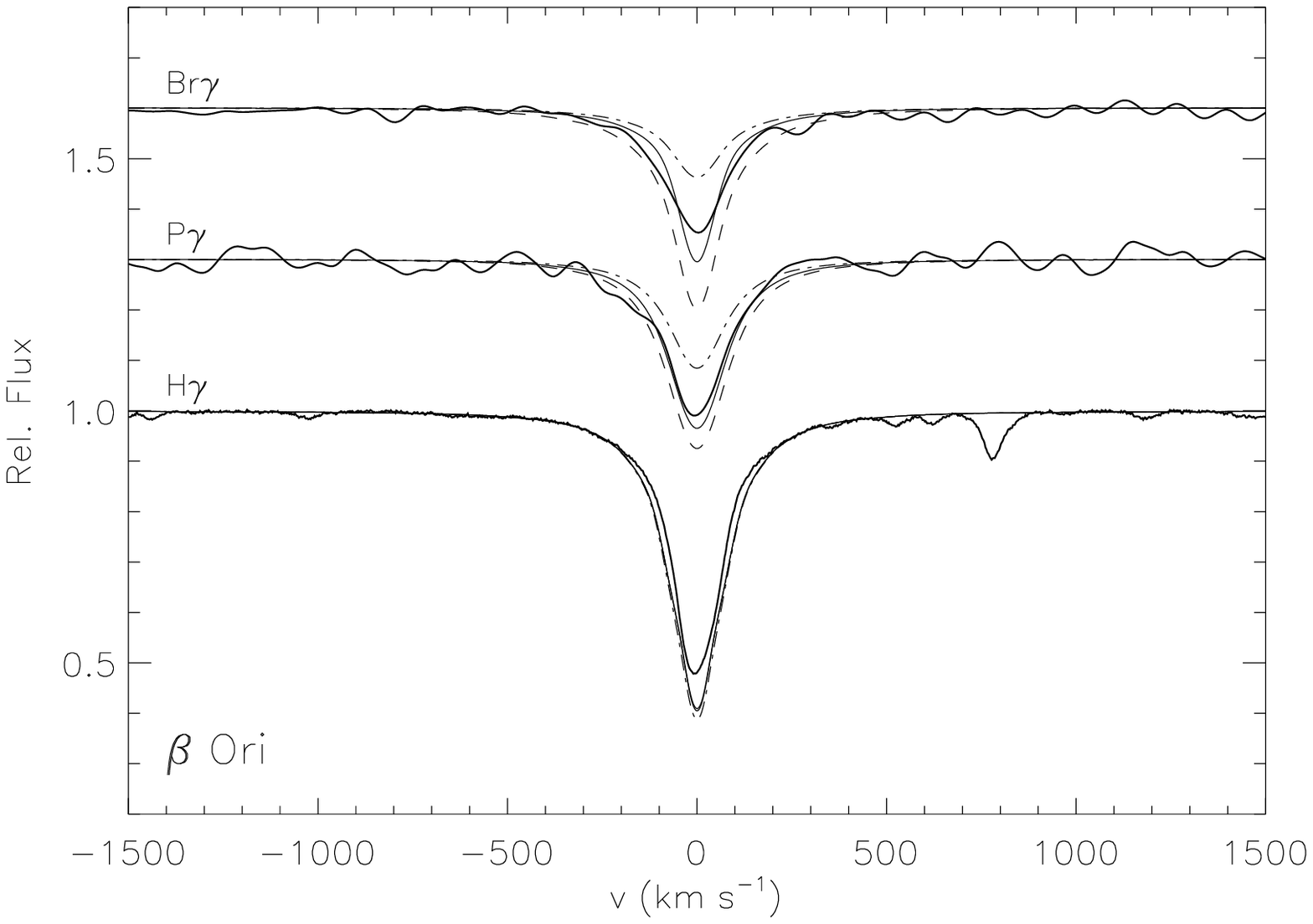}
\figcaption{
H$\gamma$, P$\gamma$ and Br$\gamma$ in $\eta$\,Leo and $\beta$\,Ori, as
indicated: models E30 ({\em full lines}), F30 ({\em dashed}) and LTE
calculations ({\em dashed-dotted}) are compared with observation ({\em thick
full lines}). 
For clarity the different spectra are shifted by 0.3 units in ordinate.
Note the lower S/N of the P$\gamma$ observations of $\eta$\,Leo.
\label{eleogamma}}
\end{figure}

\begin{figure}
\plotone{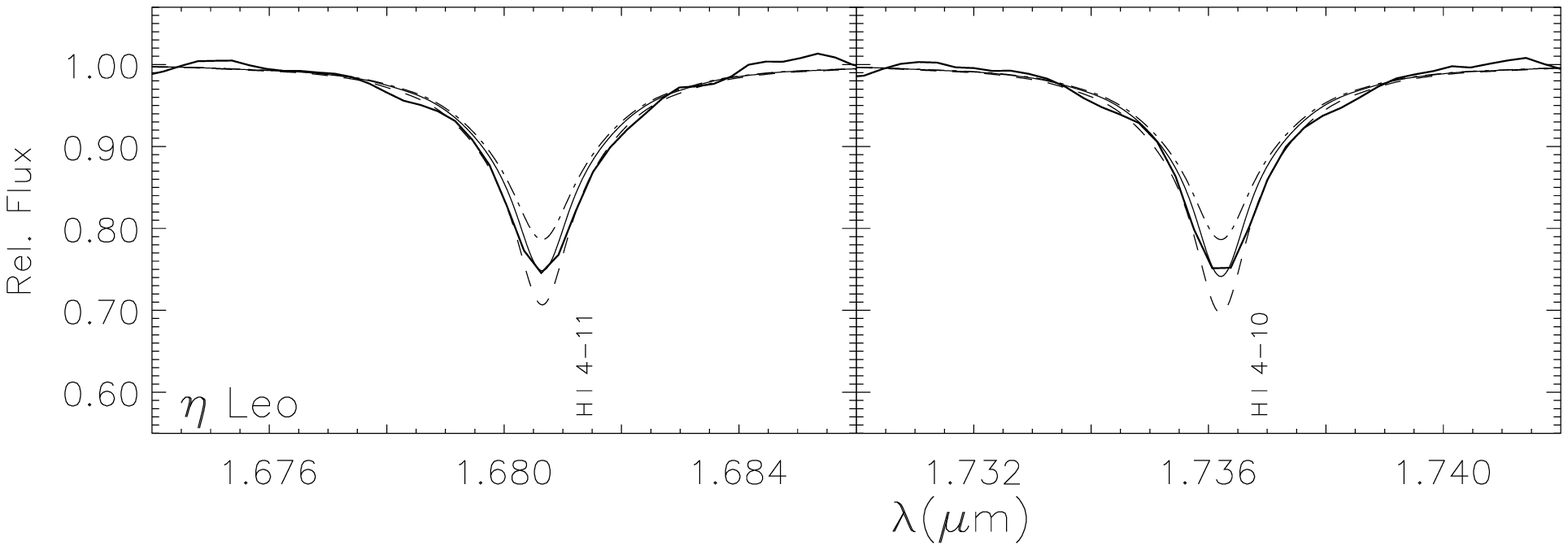}
\plotone{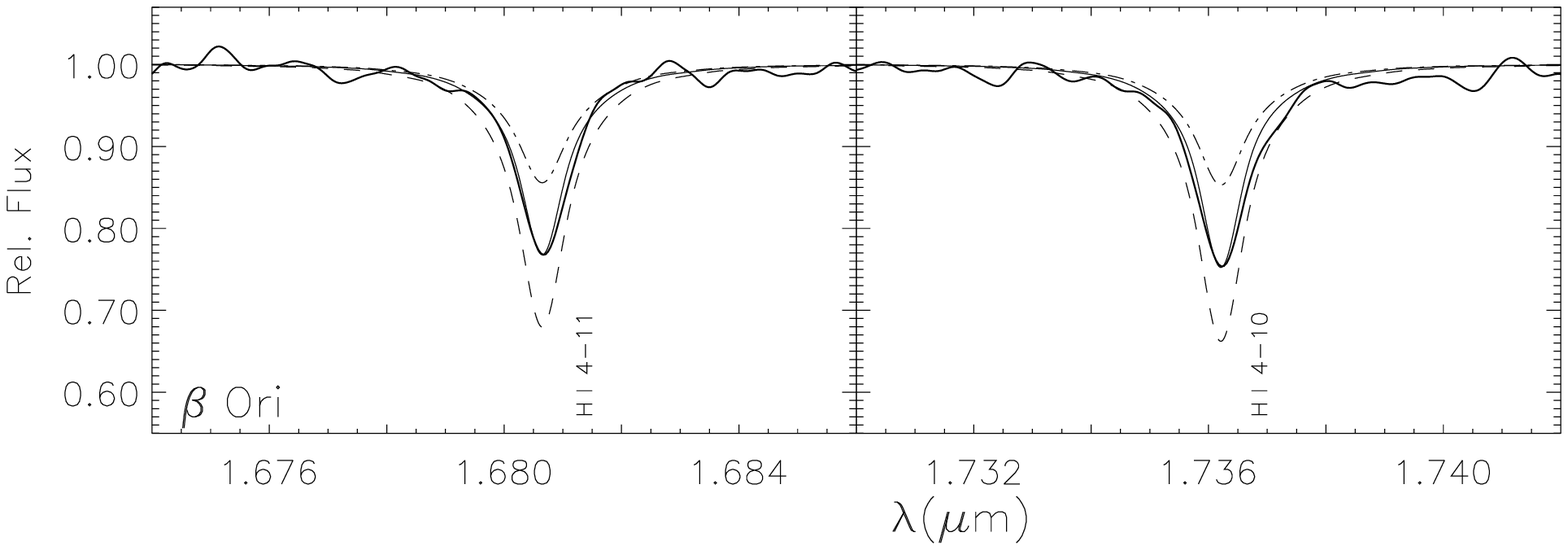}
\figcaption{
Higher Brackett lines in $\eta$\,Leo and $\beta$\,Ori, as indicated: 
models E30 ({\em full lines}), F30 ({\em dashed}) and LTE
calculations ({\em dashed-dotted}) are compared with observation ({\em thick
full lines}). 
\label{hband}}
\end{figure}

Non-LTE effects get strengthened in supergiants, and we can expect differences 
in the predictions from different model atoms to become amplified in the
same way.
Indeed, comparison of theoretical line profiles with observations for the two
bright BA-type supergiants $\eta$\,Leo and $\beta$\,Ori in Fig.~\ref{eleogamma} 
shows marked differences,
and allows us to make an important step in constraining an optimum model atom.  
H$\gamma$ in both supergiants is again barely affected by details of
the model atoms, and the deviations from LTE are small. However, stellar
wind emission may fill the line cores and consequently in such a case a good 
profile fit can only be obtained for the wings in our approach.
This also affects P$\gamma$ and Br$\gamma$, but we can definitely rule out
LTE line formation as a means of reproducing the observations, as the
predicted lines are too weak, and the
model atoms employing the J72 approximation (profiles from models B and D
resemble the model F predictions and are not shown here) appear to produce 
lines that are significantly too strong. Note that the discrepancies with P$\gamma$ in
$\eta$\,Leo may be related to the lower S/N of the spectral data. These
conclusions are confirmed by modelling of the higher Brackett lines 
in the $H$-band, which are expected to be unaffected by the stellar wind,
see Fig.~\ref{hband}. Model E gives an excellent fit, improving slightly on
model C, but all other models fail to reproduce the observations.

\begin{figure*}
\plotone{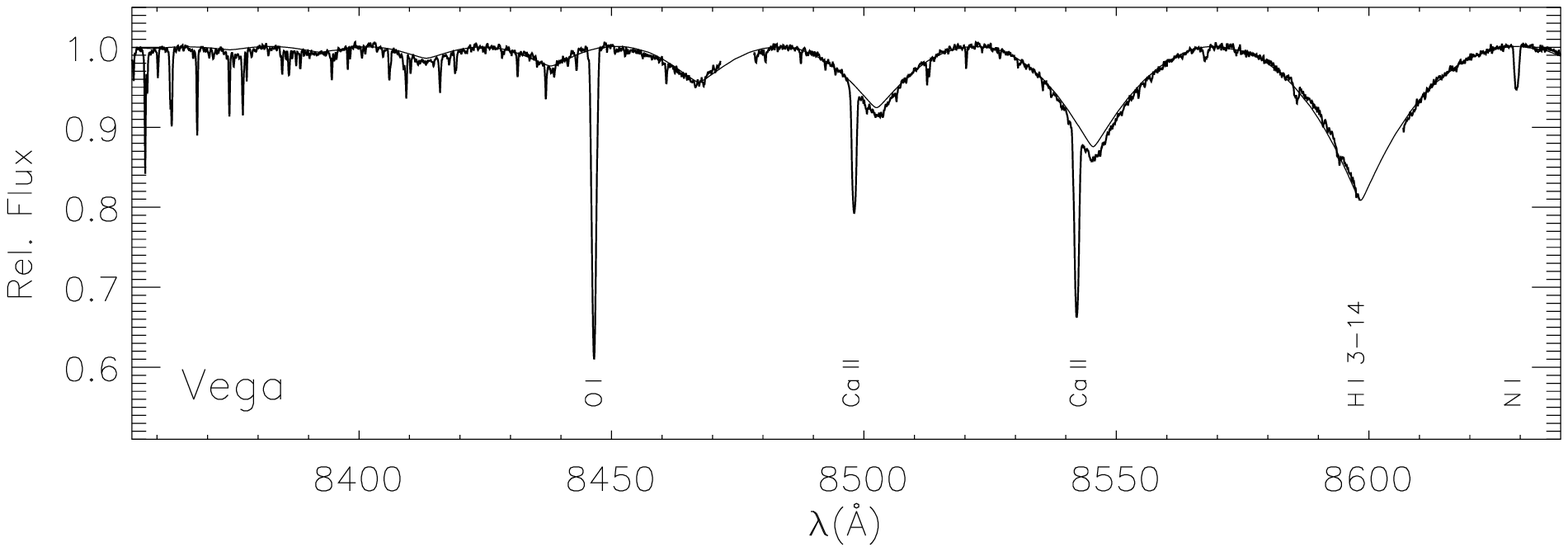}
\plotone{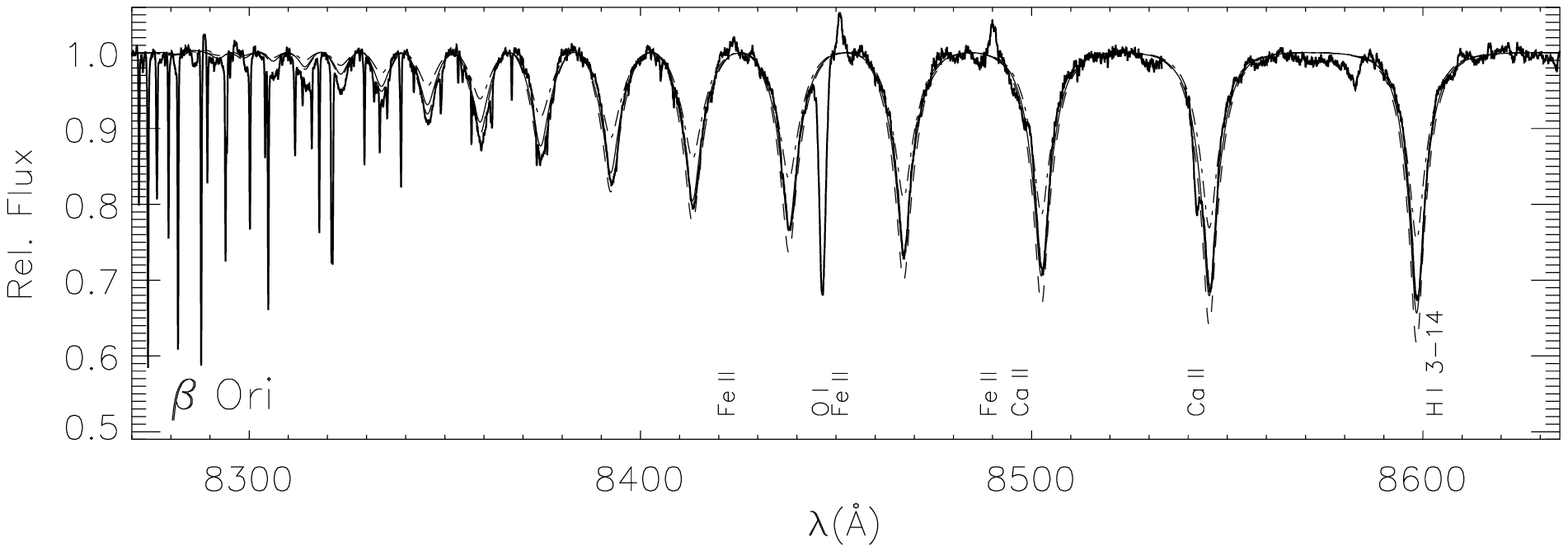}
\plotone{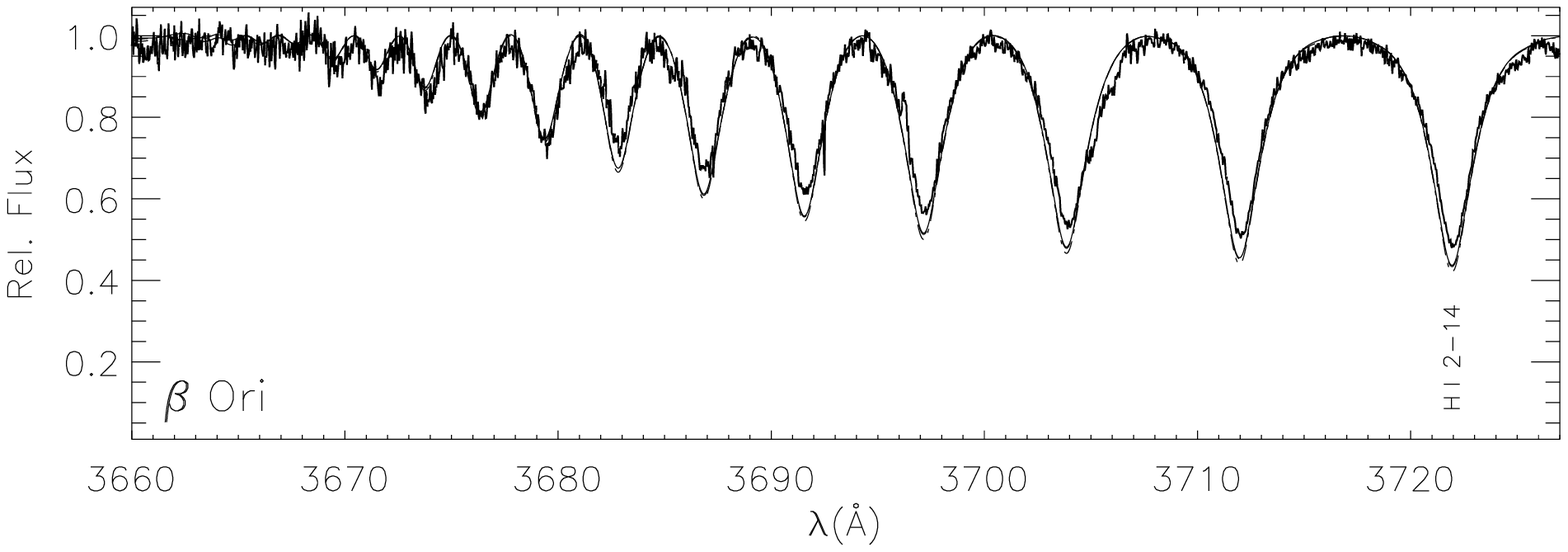}
\figcaption{
Series limits in Vega and $\beta$\,Ori, as
indicated. Models E20 ({\em full line}, top), E30 and F30 ({\em full line}
and {\em dashed}, lower two panels) and LTE profiles ({\em dashed-dotted})
are compared with observation ({\em thick full line}). Note the presence of
a few gaps in the observed spectra, and numerous sharp telluric features.
\label{pacont}}
\end{figure*}

An analogous situation is found for the lines of the Paschen series in supergiants, 
which show similar sensitivity to electron collision data as the Brackett
series members. This is in contrast to main sequence stars like Vega, where
deviations from LTE are negligible for the Paschen series, see
Fig.~\ref{pacont}. Note the high quality of the synthesis of
the series limits and the transition into the continuum, despite having not 
incorporated an occupation probability formalism into the modelling. This 
also applies to the Balmer series.

Non-LTE effects also strengthen with increasing temperature, and
despite the increasing dominance of radiative transitions over collision
processes some interesting conclusions can be drawn. In the following we will
concentrate on the main sequence, where higher atmospheric densities lead to
stronger effects of modified collision cross-sections, and lower
luminosities than in supergiants reduce the effects of the stellar wind on the 
line formation. Two stars are selected: the well-studied object $\tau$\,Sco (B0\,V)
and the most luminous main sequence object of the sample of Repolust et
al.~\cite{Repolustetal03}, HD\,93250, of spectral type O3\,V\,((f)). 

\begin{figure}
\epsscale{.75}
\plotone{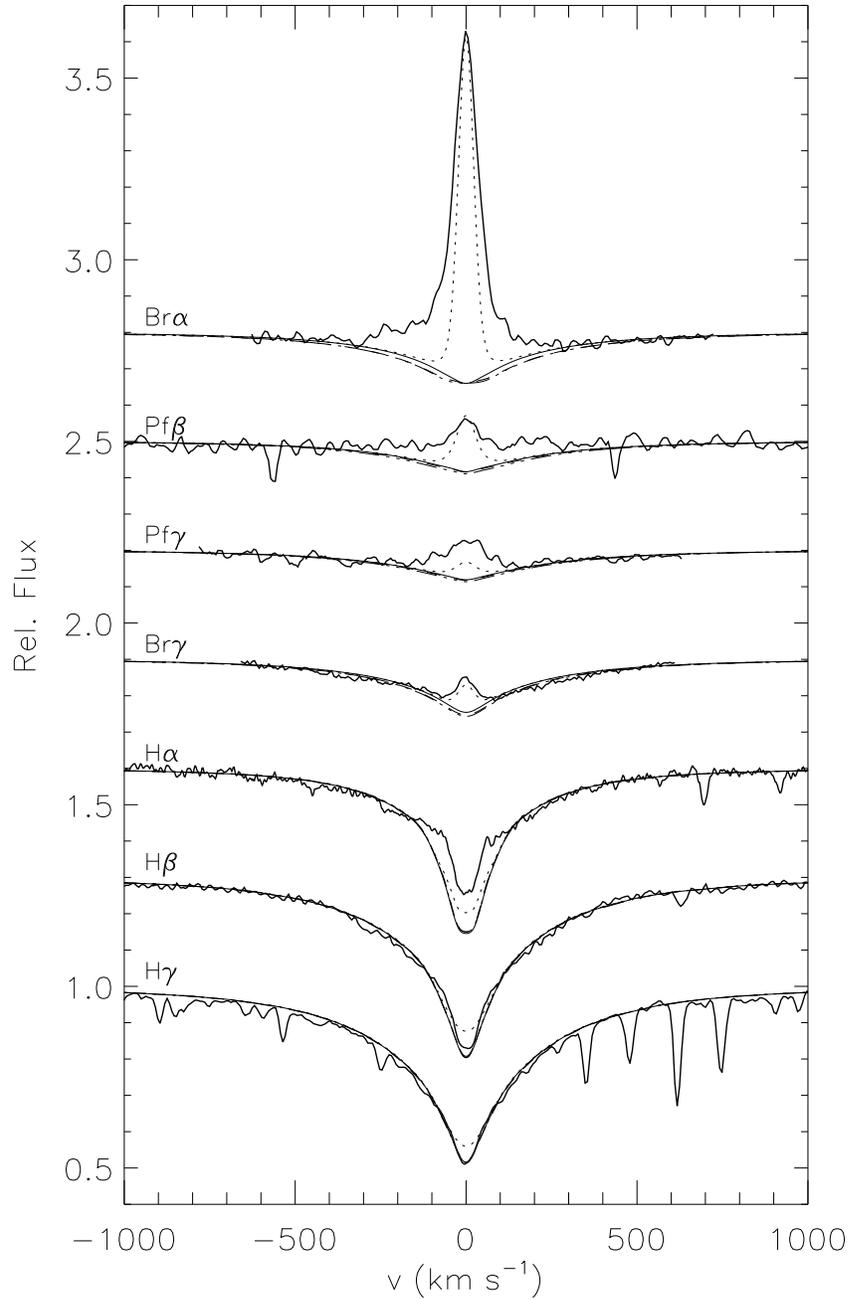}
\figcaption{Spectrum synthesis for $\tau$\,Sco -- hydrostatic/plane-parallel
modelling: models A, B, E and F ({\em dotted, dash-dot-dot-dotted, full} and
{\em dashed lines}) and observation ({\em thick full line}). All computations
are made using a 15-level model atom. For clearness the different spectra
are shifted by 0.3 units in ordinate.
\label{tausco1}}
\end{figure}

The results from our hydrostatic and plane-parallel modelling for selected
hydrogen lines of $\tau$\,Sco are displayed in Fig.~\ref{tausco1}. Detailed
collision data improves the spectrum synthesis in the very line cores of
H$\beta$ and H$\gamma$, but for H$\alpha$ and the IR-lines a good match
cannot be obtained. Intriguingly, model A provides a reasonable fit to the
IR emission profiles. We reproduce the findings of Zaal et
al.~\cite{Zaaletal99}, see their Figs.~13 and 14, {\em using a
line-blanketed LTE model atmosphere}. In view of this and our previous 
discussion we may conclude that the emission does not occur because of a
temperature inversion in the outer layers of the atmosphere, which is absent
in our LTE model atmosphere. The IR emission predicted by model A is an artifact 
of the choice of the collisional atomic data, which leads to population
inversion (see below).

The presence of a circumstellar disc around
$\tau$\,Sco may offer an alternative explanation. Assuming the star to be a 
fast rotator seen pole-on, a low-density disk may
show emission features in the IR hydrogen lines without revealing itself
via H$\alpha$-emission (Zaal, Waters \& Marlborough~1995). However,
subsequent work by Zaal et al.~\cite{Zaaletal97} finds single-peaked
IR emission features on top of broad absorption lines being
common among slow rotators rather than an exception, and they
conclude that they are an atmospheric phenomenon. This leads to suggesting
non-LTE population inversion as the source of the emission features, facilitated 
by the distinctive environment of a spherical and expanding atmosphere.

\begin{figure}
\epsscale{.75}
\plotone{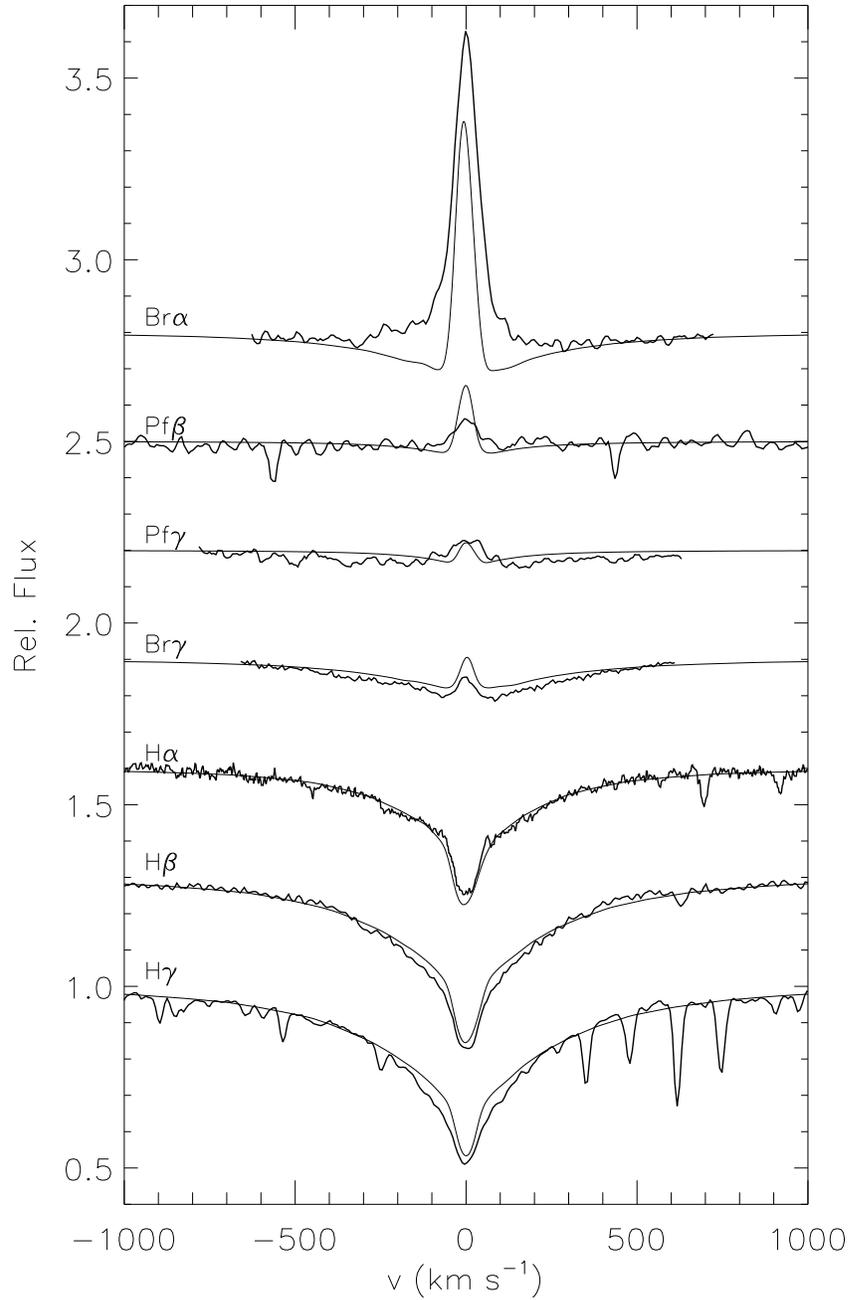}
\figcaption{Spectrum synthesis for $\tau$\,Sco -- hydrodynamic/spherical
modelling: model E15 ({\em full line}) and observation ({\em thick full
line}). A velocity field with $\beta$\,$=$\,2.4 is adopted, except for the
modelling of the Brackett lines, where we chose $\beta$\,$=$\,2.5, see the text 
for details.
\label{tausco2}}
\end{figure}

A comparison of our spectrum synthesis using a hydrodynamic approach with
observation is made in Fig.~\ref{tausco2}. Note that the observations were
obtained during four epochs spanning $\sim$10 years, so that a single set
of stellar parameters may be insufficient to reproduce all lines
simultaneously, given an observed variability of the Br$\alpha$ equivalent
width (Zaal et al.~1999). The overall agreement is
nonetheless good, with the exception of Br$\alpha$, if one allows for a slight 
adjustment of the stellar wind parameter $\beta$ (note that
$\beta$\,$\simeq$\,2.5 implies a shallower velocity law than that typical
of OB-type stars from UV/optical studies, where $\beta$\,$\simeq$\,0.8). 
We have to stress again that
even small variations of the level departures can drastically affect the
line source function in the IR (Eqn.~\ref{eqn2}) when being
close to population inversion, which turns out to be the case for $\tau$\,Sco.
In order to produce an excellent fit sound modelling is required, demanding not
only {\em perfect} atomic data but also knowledge of the {\em exact} atmospheric
conditions at line-formation depth. Several factors are limiting the
accuracy at present, most important among those the approximative treatment of the 
temperature structure and the details of connecting the
(pseudo)-hydrostatic photosphere and the stellar wind, as the
(emission) line cores are formed in the transition region.
These prohibit any further conclusions on the accuracy of the atomic data
beyond those already drawn.

Nonetheless, two additional tests have been performed for improving the 
modelling of Br$\alpha$, in particular to account for the absence of absorption 
wings in the observations, and in order to match the width (and height) of the
emission peak.
Incoherent electron scattering offers one possibility to fill
the absorption wings, via the frequency redistribution of photons from the
emission peak by an electron layer of sufficient optical depth. 
However, little improvement is achieved by accounting for this effect, as the 
stellar wind is too thin.
Clumping of the stellar wind also has the potential to affect line strengths. 
The presence of clumping in the wind of $\tau$\,Sco has been proposed by Howk et 
al.~\cite{Howketal00} in order to explain some peculiar far-UV features and the 
extremely hard X-ray emission observed in this star. In general, it is also 
supported by a 
number of UV-analyses. Based on FUSE-observations of Magellanic Cloud stars, 
Crowther et
al.~\cite{Crowtheretal02}, Massa et al.~\cite{Massaetal03} and Hillier et
al.~\cite{Hillieretal03} found indications of wind clumping.
Additional support is given by the analysis of optical data,
displaying a systematic difference of observed and theoretically predicted
wind-momentum rates for supergiants (Puls et al.~2003; Repolust et al.~2003).
Test computations for an increased mass-loss rate can reproduce the
absence of wings and the width of Br$\alpha$, however without properly matching
the height of the emission peak simultaneously. The $\dot{M}$ required for
achieving this is approximately an order of magnitude higher than the
present value, bringing the predictions for the other hydrogen lines into
disagreement with observation. We thus cannot rule out clumping effects
in the case of $\tau$\,Sco, however a more sophisticated approach should be 
investigated separately, as this goes beyond the scope of the present study.

\begin{figure}
\plotone{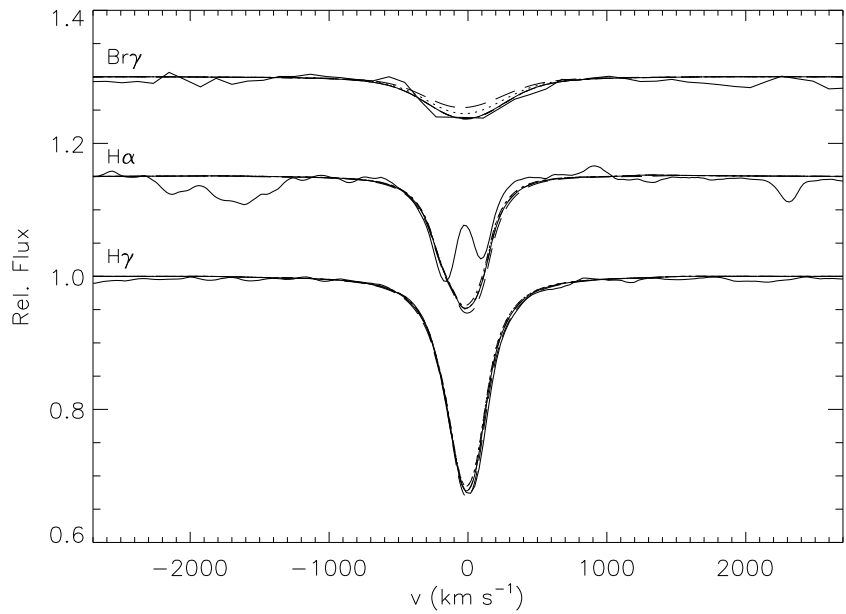}
\figcaption{Spectrum synthesis for HD\,93250 -- hydrodynamic/spherical
modelling: models A, B, E, F and X ({\em dotted, dash-dot-dot-dotted, full,
dashed} and {\em long dashed lines}) and observation ({\em thick full line}).
All computations are performed using a 20-level model atom. Note the nebular
emission component in the H$\alpha$ core. For clarity the
different spectra are shifted by 0.15 units in ordinate.
\label{HD93250}}
\end{figure}

Finally, HD\,93250 may act as a benchmark for the study of objects at the earliest
phases of stellar evolution of massive stars. Typically, the very young
objects are deeply embedded in a (ultra)-compact \ion{H}{2}-region
prohibiting spectroscopy in the visual. In order to study the immediate outcome
of the star-formation process, stellar parameters have to be determined
from the IR features. We conclude from Fig.~\ref{HD93250} that simultaneous 
agreement between the modelling in the visual and in the $K$-band is
achieved if detailed electron collision data is accounted for (as in
model E). Here, radiative transitions dominate over collisions, such
that most of the model profiles coincide. The only noticeable differences
occur for model A, where the predicted equivalent widths are smaller by
$\sim$15\%. Computations
using model atom X have been performed in this case only. In addition to the
inconsistencies in the GNP dataset from theoretical considerations, as 
reported by Chang, Avrett \& Loeser~\cite{CAL91}, the computations indicate 
a significantly poorer fit than achieved using model atoms A
(Fig.~\ref{HD93250}). Consequently, these data should be
avoided in quantitative non-LTE modelling of the hydrogen IR lines. 

\section{Recommendations}
In view of the comparison of the predictions of different model atoms
for \ion{H}{1} with observation we can conclude that the use of electron
collision data from {\em ab-initio} computations is {\em mandatory} in order
to derive consistent results for the visual and IR. Model atoms relying
on the MHA and J72 approximation data obviously fail in achieving this over the 
entire range of OBA-type stars. Introduction of
detailed collision data such as those of ABBS or B04 largely remove the
discrepancies. However, slight differences in the atomic data in combination 
with the use of approximation data for transitions involving levels with
high $n$, which cannot be entirely abandoned in the modelling, still result in  
a variety of distinguishable predictions. Among the model atoms implemented, 
model E provides the best overall agreement between observation and the 
spectrum synthesis over the whole parameter space investigated here.
Consequently, we recommend the data of B04 for the evaluation of collision rates of
\ion{H}{1}, supplemented by the approximation formulae by PR and of MHA for those 
transitions not covered by the {\em ab-initio} computations. This applies
not only to the modelling of massive stars as we have done but to all
hydrogen plasmas. 

With regard to the model complexity for the non-LTE computations we suggest the use of a 
model atom with a number of explicitly treated levels according to the
classical Inglis-Teller limit for a given star, when not accounting for the
occupation probability formalism as implemented by Hubeny et
al.~\cite{HHL94}. Use of the atomic data recommended above further helps to
minimise artifacts introduced by a model atom of inappropriate complexity.

\acknowledgments
We are grateful to J.~Puls for providing the {\sc Fastwind} code and valuable
comments on the manuscript, and to F.~Najarro and R.P.~Kudritzki for 
helpful discussion. 
This work benefited much from the kind provision of spectra by A.~Fullerton,
K. Hinkle
and P.A.~Zaal. The NSO/Kitt Peak FTS data used here were produced by NSF/NOAO.

\appendix
\section{Electron-impact excitation data}
Here we summarise some results of the {\em ab-initio} computations of Butler
(2004, in preparation) for electron-impact excitation cross-sections in
hydrogen relevant for the present work. The $R$-matrix method in the
close-coupling approximation was used to obtain data for hydrogen for all transitions
between states with principal quantum number $n$\,$\leq$\,7 and angular
momentum quantum number $\ell$\,$\leq$\,6. For practical applications 
thermally-averaged effective collision strengths between lower 
and upper states $i$ and $j$ may be defined
\begin{equation}
\Upsilon_{ij}=\int_{0}^{\infty} \Omega_{ij} \exp(E_j/kT)
d(E_j/kT)
\end{equation}
where $\Omega_{ij}$ is the collision strength and $E_j$ the kinetic energy of the 
outgoing electron. The excitation rate coefficient is then
\begin{equation}
q_{ij}=\frac{8.63 \times 10^{-6}}{g_i T^{1/2}} \Upsilon_{ij}
\exp(\Delta E_{ij}/kT)\,{\rm cm^3\,s^{-1}}
\end{equation}
where $g_i$ is the statistical weight of state $i$ and $\Delta E_{ij}$ the energy
difference between the two states. The effective collision strength is
dimensionless and symmetric, i.e. $\Upsilon_{ij}$\,$=$\,$\Upsilon_{ji}$.

Data for transitions between levels $n$ and $n'$, i.e. summarised over the
degenerate states with different quantum number $\ell$, are given in  
Table~\ref{upsilontab} for a wide range of temperatures. Details on the 
computations and the entire dataset will be 
discussed by Butler (2004, in preparation).

\section{Gaunt factors}  
  The gaunt factors that we use (Wright,~1978) have not been published so we 
give the relevant data here.  The fits have the form
\begin{eqnarray}
g_{\rm II}&=&a_1+\frac{b_1}{\nu}+\frac{c_1}{\nu^2}\qquad 10^{16}\,{\rm s}^{-1} > \nu >
\frac{3.28805\times10^{15}}{n^2}\,{\rm s}^{-1}\nonumber \\
g_{\rm II}&=&a_2+\frac{b_2}{\nu}+\frac{c_2}{\nu^2}\qquad 6\times
10^{16}\,{\rm s}^{-1}>\nu>10^{16}\,{\rm s}^{-1}\nonumber
\end{eqnarray}
and the coefficients are to be found in Table~\ref{tabgaunt}. The
uncertainties of the fits are typically less than 0.3\%.

\clearpage

\begin{deluxetable}{ll}
\tablecaption{Non-LTE Model Atoms\label{tabmod}}
\tablehead{\colhead{Model} & \colhead{Electron-impact excitation data}}
\startdata
A & MHA (all $n$,$n'$)\\
B & J72 (all $n$,$n'$)\\
C & ABBS ($n$,$n'$\,$\leq$\,5), MHA (rest)\\
D & ABBS ($n$,$n'$\,$\leq$\,5), J72 (rest)\\
E & B04 ($n$,$n'$\,$\leq$\,7), PR ($n$,$n'$\,$\ge$\,5), MHA (rest)\\
F & B04 ($n$,$n'$\,$\leq$\,7), PR ($n$,$n'$\,$\ge$\,5), J72 (rest)\\
X & GNP ($n$,$n'$\,$\leq$\,15), MHA (rest)
\enddata
\tablecomments{see text for references}
\end{deluxetable}

\begin{deluxetable}{llrrrrrrrrrr}
\tablecaption{Stellar Parameters\label{tabpara}}
\tablehead{Object & \colhead{Sp.\,Type} & \colhead{$T_{\rm eff}$} &
\colhead{$\log g$} & \colhead{$y$} & \colhead{$R$/R$_{\odot}$} &
\colhead{$\log L$/L$_{\odot}$} & \colhead{$v_{\rm t}$} & 
\colhead{$v\,\sin i$} & \colhead{$\dot{M}$} &
\colhead{$v_{\infty}$} & \colhead{$\beta$}\\
 & & (K) & (cgs) & & & & (km\,s$^{-1}$) & (km\,s$^{-1}$) &
(10$^{-6}$\,M$_{\odot}$\,yr$^{-1}$) & (km\,s$^{-1}$)}
\tabletypesize{\scriptsize}
\rotate
\tablewidth{0pt}
\startdata
\noindent
Vega            & A0\,V  &  9550 & 3.95 & 0.09  & 2.8  & 1.78 & 2 & 22  & {\ldots} & {\ldots} & {\ldots}\\
$\tau$\,Sco & B0\,V  & 31400 & 4.24 & 0.09  & 5.1  & 4.36 & 3 & 19  & 0.009 & 2000 & 2.4/2.5\\
\objectname{HD93250} & O3.5\,V  & 46000 & 3.95 & 0.09  & 15.9 & 6.01 & 0 & 130 & 3.45 & 3250 & 0.9\\
$\eta$\,Leo     & A0\,Ib &  9600 & 2.00 & 0.13  &  50  & 4.28 & 4 &  9  & {\ldots} & {\ldots} & {\ldots}\\
$\beta$\,Ori    & B8\,Ia & 12000 & 1.75 & 0.135 & 104  & 5.30 & 7 & 36  & {\ldots} & {\ldots} & {\ldots}
\enddata
\end{deluxetable}

\clearpage

\begin{table}
\scriptsize
\caption{Effective collision strengths $\Upsilon_{nn'}$ for the 
$n \longrightarrow n'$ transitions in H
\label{upsilontab}}
\begin{tabular}{llllllllll}
\tableline
\tableline
& & \multicolumn{8}{c}{$T$\,(K)}\\
\cline{3-10}
$n$ & $n'$ & 2500 & 5000 & 7500 & 10000 & 15000 & 20000 & 25000 & 30000\\
\tableline
1 & 2 &	6.40($-$1)\tablenotemark{a} & 6.98($-$1) & 7.57($-$1) & 8.09($-$1) & 8.97($-$1) & 9.78($-$1) & 1.06($+$0) & 1.15($+$0)\\
1 & 3 &	2.20($-$1) & 2.40($-$1) & 2.50($-$1) & 2.61($-$1) & 2.88($-$1) & 3.22($-$1) & 3.59($-$1) & 3.96($-$1)\\
1 & 4 &	9.93($-$2) & 1.02($-$1) & 1.10($-$1) & 1.22($-$1) & 1.51($-$1) & 1.80($-$1) & 2.06($-$1) & 2.28($-$1)\\
1 & 5 &	4.92($-$2) & 5.84($-$2) & 7.17($-$2) & 8.58($-$2) & 1.12($-$1) & 1.33($-$1) & 1.50($-$1) & 1.64($-$1)\\
1 & 6 &	2.97($-$2) & 4.66($-$2) & 6.28($-$2) & 7.68($-$2) & 9.82($-$2) & 1.14($-$1) & 1.25($-$1) & 1.33($-$1)\\
1 & 7 &	5.03($-$2) & 6.72($-$2) & 7.86($-$2) & 8.74($-$2) & 1.00($-$1) & 1.10($-$1) & 1.16($-$1) & 1.21($-$1)\\
2 & 3 &	2.35($+$1) & 2.78($+$1) & 3.09($+$1) & 3.38($+$1) & 4.01($+$1) & 4.71($+$1) & 5.45($+$1) & 6.20($+$1)\\
2 & 4 &	1.07($+$1) & 1.15($+$1) & 1.23($+$1) & 1.34($+$1) & 1.62($+$1) & 1.90($+$1) & 2.18($+$1) & 2.44($+$1)\\
2 & 5 &	5.22($+$0) & 5.90($+$0) & 6.96($+$0) & 8.15($+$0) & 1.04($+$1) & 1.23($+$1) & 1.39($+$1) & 1.52($+$1)\\
2 & 6 &	2.91($+$0) & 4.53($+$0) & 6.06($+$0) & 7.32($+$0) & 9.17($+$0) & 1.05($+$1) & 1.14($+$1) & 1.21($+$1)\\
2 & 7 &	5.25($+$0) & 7.26($+$0) & 8.47($+$0) & 9.27($+$0) & 1.03($+$1) & 1.08($+$1) & 1.12($+$1) & 1.14($+$1)\\
3 & 4 &	1.50($+$2) & 1.90($+$2) & 2.28($+$2) & 2.70($+$2) & 3.64($+$2) & 4.66($+$2) & 5.70($+$2) & 6.72($+$2)\\
3 & 5 &	7.89($+$1) & 9.01($+$1) & 1.07($+$2) & 1.26($+$2) & 1.66($+$2) & 2.03($+$2) & 2.37($+$2) & 2.68($+$2)\\
3 & 6 &	4.13($+$1) & 6.11($+$1) & 8.21($+$1) & 1.01($+$2) & 1.31($+$2) & 1.54($+$2) & 1.72($+$2) & 1.86($+$2)\\
3 & 7 &	7.60($+$1) & 1.07($+$2) & 1.25($+$2) & 1.37($+$2) & 1.52($+$2) & 1.61($+$2) & 1.68($+$2) & 1.72($+$2)\\
4 & 5 &	5.90($+$2) & 8.17($+$2) & 1.07($+$3) & 1.35($+$3) & 1.93($+$3) & 2.47($+$3) & 2.96($+$3) & 3.40($+$3)\\
4 & 6 &	2.94($+$2) & 4.21($+$2) & 5.78($+$2) & 7.36($+$2) & 1.02($+$3) & 1.26($+$3) & 1.46($+$3) & 1.64($+$3)\\
4 & 7 &	4.79($+$2) & 7.06($+$2) & 8.56($+$2) & 9.66($+$2) & 1.11($+$3) & 1.21($+$3) & 1.29($+$3) & 1.34($+$3)\\
5 & 6 &	1.93($+$3) & 2.91($+$3) & 4.00($+$3) & 5.04($+$3) & 6.81($+$3) & 8.20($+$3) & 9.29($+$3) & 1.02($+$4)\\
5 & 7 &	1.95($+$3) & 3.24($+$3) & 4.20($+$3) & 4.95($+$3) & 6.02($+$3) & 6.76($+$3) & 7.29($+$3) & 7.70($+$3)\\
6 & 7 &	6.81($+$3) & 1.17($+$4) & 1.50($+$4) & 1.73($+$4) & 2.03($+$4) & 2.21($+$4) & 2.33($+$4) & 2.41($+$4)\\
\tableline
\end{tabular}
\vspace{.5cm}
\tablenotetext{a}{$a(b)$: $a\times 10^{b}$}
\begin{tabular}{llllllllll}
\tableline
\tableline
& & \multicolumn{8}{c}{$T$\,(K)}\\
\cline{3-10}
$n$ & $n'$ & 40000 & 50000 & 60000 & 80000 & 100000 & 150000 & 200000 & 250000\\
\tableline
1 & 2 &  1.32($+$0) & 1.51($+$0) & 1.68($+$0) & 2.02($+$0) & 2.33($+$0) & 2.97($+$0) & 3.50($+$0) & 3.95($+$0)\\
1 & 3 &	 4.64($-$1) & 5.26($-$1) & 5.79($-$1) & 6.70($-$1) & 7.43($-$1) & 8.80($-$1) & 9.79($-$1) & 1.06($+$0)\\
1 & 4 &	 2.66($-$1) & 2.95($-$1) & 3.18($-$1) & 3.55($-$1) & 3.83($-$1) & 4.30($-$1) & 4.63($-$1) & 4.88($-$1)\\
1 & 5 &	 1.85($-$1) & 2.01($-$1) & 2.12($-$1) & 2.29($-$1) & 2.39($-$1) & 2.59($-$1) & 2.71($-$1) & 2.81($-$1)\\
1 & 6 &	 1.45($-$1) & 1.53($-$1) & 1.58($-$1) & 1.65($-$1) & 1.70($-$1) & 1.77($-$1) & 1.82($-$1) & 1.85($-$1)\\
1 & 7 &	 1.27($-$1) & 1.31($-$1) & 1.34($-$1) & 1.36($-$1) & 1.37($-$1) & 1.39($-$1) & 1.39($-$1) & 1.40($-$1)\\
2 & 3 &	 7.71($+$1) & 9.14($+$1) & 1.05($+$2) & 1.29($+$2) & 1.51($+$2) & 1.93($+$2) & 2.26($+$2) & 2.52($+$2)\\
2 & 4 &	 2.89($+$1) & 3.27($+$1) & 3.60($+$1) & 4.14($+$1) & 4.56($+$1) & 5.31($+$1) & 5.83($+$1) & 6.23($+$1)\\
2 & 5 &	 1.74($+$1) & 1.90($+$1) & 2.03($+$1) & 2.23($+$1) & 2.37($+$1) & 2.61($+$1) & 2.78($+$1) & 2.89($+$1)\\
2 & 6 &	 1.31($+$1) & 1.38($+$1) & 1.44($+$1) & 1.51($+$1) & 1.56($+$1) & 1.63($+$1) & 1.68($+$1) & 1.71($+$1)\\
2 & 7 &	 1.17($+$1) & 1.18($+$1) & 1.19($+$1) & 1.19($+$1) & 1.20($+$1) & 1.19($+$1) & 1.19($+$1) & 1.19($+$1)\\
3 & 4 &	 8.66($+$2) & 1.04($+$3) & 1.19($+$3) & 1.46($+$3) & 1.67($+$3) & 2.08($+$3) & 2.39($+$3) & 2.62($+$3)\\
3 & 5 &	 3.19($+$2) & 3.62($+$2) & 3.98($+$2) & 4.53($+$2) & 4.95($+$2) & 5.68($+$2) & 6.16($+$2) & 6.51($+$2)\\
3 & 6 &	 2.08($+$2) & 2.24($+$2) & 2.36($+$2) & 2.53($+$2) & 2.65($+$2) & 2.83($+$2) & 2.94($+$2) & 3.02($+$2)\\
3 & 7 &	 1.78($+$2) & 1.81($+$2) & 1.83($+$2) & 1.85($+$2) & 1.86($+$2) & 1.87($+$2) & 1.86($+$2) & 1.87($+$2)\\
4 & 5 &	 4.14($+$3) & 4.75($+$3) & 5.25($+$3) & 6.08($+$3) & 6.76($+$3) & 8.08($+$3) & 9.13($+$3) & 1.00($+$4)\\
4 & 6 &	 1.92($+$3) & 2.15($+$3) & 2.33($+$3) & 2.61($+$3) & 2.81($+$3) & 3.15($+$3) & 3.36($+$3) & 3.51($+$3)\\
4 & 7 &	 1.41($+$3) & 1.46($+$3) & 1.50($+$3) & 1.55($+$3) & 1.57($+$3) & 1.61($+$3) & 1.62($+$3) & 1.63($+$3)\\
5 & 6 &	 1.15($+$4) & 1.26($+$4) & 1.34($+$4) & 1.49($+$4) & 1.63($+$4) & 1.97($+$4) & 2.27($+$4) & 2.54($+$4)\\
5 & 7 &	 8.26($+$3) & 8.63($+$3) & 8.88($+$3) & 9.21($+$3) & 9.43($+$3) & 9.78($+$3) & 1.00($+$4) & 1.02($+$4)\\
6 & 7 &	 2.52($+$4) & 2.60($+$4) & 2.69($+$4) & 2.90($+$4) & 3.17($+$4) & 3.94($+$4) & 4.73($+$4) & 5.50($+$4)\\
\tableline
\end{tabular}
\end{table}

\begin{deluxetable}{cccrccc}
\tablecaption{Coefficients of polynomial fits to the bound-free gaunt factors
\label{tabgaunt}}
\tablehead{\colhead{$n$} & \colhead{$a_1$} & \colhead{$b_1$} & 
\colhead{$c_1$} & \colhead{$a_2$} & \colhead{$b_2$} & \colhead{$c_2$}} 
\startdata
1 & 1.0780 & $-$8.754(14)\tablenotemark{a} & $-$1.791(29) & 0.798 & 5.358(15) & $-$3.484(31) \\
2 & 1.0926 & $-$2.019(14) &  1.836(28) & 0.768 & 6.242(15) & $-$3.208(31) \\
3 & 1.0983 & $-$9.450(13) &  9.177(27) & 0.793 & 5.480(15) & $-$2.318(31) \\
4 & 1.0954 & $-$5.188(13) &  3.552(27) & 0.831 & 4.094(15) & $-$1.430(31) \\
5 & 1.0912 & $-$3.200(13) &  1.576(27) & 0.758 & 6.633(15) & $-$3.320(31) \\
6 & 1.0925 & $-$2.331(13) &  9.325(26) & 0.790 & 5.808(15) & $-$2.844(31) 
\enddata
\tablecomments{$a(b)$: $a\times 10^{b}$}
\end{deluxetable}

\end{document}